\documentclass[11pt,preprint]{aastex}
\usepackage{lscape}

\def \etal{{et~al.\null}}

\def\lea{\mathrel{<\kern-1.0em\lower0.9ex\hbox{$\sim$}}}
\def\gea{\mathrel{>\kern-1.0em\lower0.9ex\hbox{$\sim$}}}
\newcommand{\lta}{{\>\rlap{\raise2pt\hbox{$<$}}\lower3pt\hbox{$\sim$}\>}}
\newcommand{\gta}{{\>\rlap{\raise2pt\hbox{$>$}}\lower3pt\hbox{$\sim$}\>}}

\begin{document}

\shorttitle{ALMA Observations of the Antennae Galaxies}

\title{ALMA Observations of the Antennae Galaxies: I. A New Window on a Prototypical Merger}

\author{
Bradley C.\ Whitmore,\altaffilmark{1}
Crystal Brogan,\altaffilmark{2}
Rupali Chandar,\altaffilmark{3} 
Aaron Evans,\altaffilmark{2,4}
John Hibbard,\altaffilmark{2}
Kelsey Johnson,\altaffilmark{4,5}
Adam Leroy,\altaffilmark{2}
George Privon,\altaffilmark{4}
Anthony Remijan,\altaffilmark{2} and
Kartik Sheth\altaffilmark{2}
}

\affil{email: whitmore@stsci.edu}

\altaffiltext{1}{Space Telescope Science Institute, Baltimore, MD, USA, 21218}
\altaffiltext{2}{National Radio Astrophysical Observatory, Charlottesville, VA, 22903, USA}
\altaffiltext{3}{Department of Physics \& Astronomy, The University of Toledo, Toledo, OH, 43606, USA}
\altaffiltext{4}{Department of Astronomy, University of Virginia, Charlottesville, VA 22904, USA}
\altaffiltext{5}{Adjunct Astronomer at the National Radio Astrophysical Observatory, Charlottesville, VA, 22903, USA}

\begin{abstract}
We present the highest spatial resolution ($\approx$0.5$''$) CO (3-2) observations to date of the ``overlap" region in the merging Antennae galaxies (NGC~4038/39), taken with the Atacama Large Millimeter/sub-millimeter Array (ALMA).  We report on the  discovery of 
a long  (3~kpc), thin (aspect ratio 30/1), filament of CO gas which breaks up into roughly ten individual knots. Each individual knot has a low internal velocity dispersion ($\approx$10~km~s$^{-1}$), and the dispersion of the ensemble of knots in the filament is also low ($\approx$10~km~s$^{-1}$).  
At the other extreme, we find that the individual clouds in the Super Giant Molecular Cloud~2 region discussed by Wilson and collaborators  have a large range of internal velocity dispersions (10 to 80~km~s$^{-1}$), and a large dispersion amongst the ensemble ($\approx$80~km~s$^{-1}$). Other large-scale features observed in CO emission, and their correspondence with historical counterparts using observations in other  wavelengths, are also discussed.  We compare the locations of small-scale CO features with a variety 
of multi-wavelength observations, in particular broad- (\textit{BVIJH}) and narrow-band data (H$_{\alpha}$ and Pa$_{\beta}$) taken with 
\textit{HST}, and radio (3.6~cm) continuum observations taken with the Karl~G.\  Jansky Very Large Array (VLA). This comparison lead to the development of an evolutionary classification system which provides a framework for studying the sequence of star cluster formation and evolution, from diffuse SGMCs, to proto, embedded, emerging, young, intermediate/old clusters.  The relative timescales have been assessed by determining the fractional population of sources at each evolutionary stage. The main uncertainty in this estimate is the identification of four regions as candidate protoclusters (i.e., strong compact CO emission  but no clearly associated radio emission).  Using the evolutionary framework, we estimate the maximum age range of clusters in a single SGMC is $\approx$10~Myr,  which suggests that the molecular gas is removed over this timescale resulting in the cessation of star formation and the destruction of the SGMC within a radius of about 200~pc.
\end{abstract}

\keywords{galaxies: individual (Antennae, NGC 4038/39) --- galaxies: star clusters --- stars: formation}

\section{Introduction}

The Antennae galaxies (NGC~4038/39) first drew the attention of Duncan (1923) because of their peculiar shape  (Figure~1). It was not until the 1970s that it became clear that this object was a pair of colliding galaxies (Rubin, Ford, D'Odorico 1970; Toomre \& Toomre 1972). The system subsequently gained prominence as one of the youngest and closest examples of a galactic merger in the famous Toomre (1977) sequence.  With the discovery that essentially all strong extragalactic  IR sources are mergers (Sanders et~al.\ 1988), the Antennae began to attract attention as an ideal laboratory for the study of star formation, especially in the form of super star clusters (e.g., Whitmore \& Schweizer 1995; Mirabel et~al.\ 1998; Wilson et~al.\ 2000).

The Antennae have an infrared luminosity LIR~= $10^{11}\,L_{\odot}$, hence they qualify as a luminous infrared galaxy, though just barely.  In addition, CO mapping reveals a molecular gas reservoir that is vast, even considering the amount of on-going star formation in the Antennae ($1.5\times 10^{10}\,M_{\odot}$; Gao et~al.\  2001, Wilson et~al.\ 2003), suggesting that the Antennae may evolve into an ultraluminous infrared galaxy (ULIRG) on a fairly short timescale (e.g., Gao et~al.\ 2001; Karl et~al.\ 2010).  

Until recently, the large difference in spatial resolution between optical observations (0.1$''$) and longer wavelength observations (e.g., $3\times5''$ for CO observations by Wilson 2000; 2$''$ for observations with the \textit{Spitzer Space Telescope} by Brandl et~al.\ 2009; 1--1.5$''$ for SMA observations by Ueda et~al.\ 2012 and Wei et~al.\ 2012) have been a limiting factor in the study of the Antennae. Similarly, the available velocity resolution for SGMCs in the Antennae provided limited kinematic information.

The Antennae were first observed by ALMA (Atacama Large
Millimeter/submillimeter Array) in the science verification (SV) period in both CO (2-1) and CO (3-2) with resolutions of $1.68 ''  \times 0.85 ''$ and $1.1''  \times 0.6 ''$, respectively.  The SV CO (2-1) data revealed a previously unidentified molecular gas arm (Espada et~al.\ 2012), and the SV CO (3-2) data were able to resolve molecular structures at a level not previously possible (Herrera et~al.\ 2012). In particular, Herrera et~al.\ (2011, 2012) note the presence of a luminous and compact CO source in the ``overlap'' region of the
Antennae that has strong H$_2$ emission associated with it, and they
hypothesize that it may be a progenitor to a super star cluster.  The new Cycle~0 ALMA observations reported here, with $\approx$0.5$''$ (i.e., $0.56''\times0.43''$ synthesized beam size) spatial resolution and $\approx$5~km~s$^{-1}$ velocity resolution for moderate S/N features  represent a major improvement in our ability to study the earliest stages of star cluster formation in the Antennae.\footnote{Note that 0.5~km~s$^{-1}$ velocity resolution is possible for the highest S/N regions.}

In recent years, the Antennae have also become one of the primary
``proving grounds'' for many new telescopes and instruments, due both
to the relative proximity of the galaxy and the brightness of the molecular gas and super star clusters.  One example of particular relevance for this work is multi-band imaging taken with the \textit{HST} using four generations
of cameras (Whitmore et~al.\ 1995, 1999, 2010). We have used these
observations to estimate the age, extinction, and mass for many thousands of star clusters.  In the current paper we continue this tradition of observing the Antennae early in the lifetime of a new telescope by reporting on Cycle~0 observations using ALMA.

The goal of this paper is to present the data and to briefly examine both large- and small-scale features in the Cycle~0 ALMA CO~(3-2) observations.
The paper also acts as a background for more detailed papers to follow.  These include a survey of the CO clouds in the Antennae and their associated ``scaling laws'' (Paper~2; Leroy et~al.\ 2014, in preparation), an
examination of the ALMA~850~$\mu$m continuum maps and the dense gas
tracers HCN~(4-3) and HCO$^{+}$~(4-3) (Paper~3; Brogan et~al.\ 2014, in preparation), a detailed look at the ``Firecracker,'' a proto super star
cluster in Super Giant Molecular Cloud 2 (i.e., SGMC2; Paper~4; Johnson 
et~al.\ 2014), and further elaboration of the multi-wavelength evolutionary cluster classification system discussed in the current paper (Paper~5; Whitmore et~al.\ 2015, in preparation).

In this paper we adopt a distance modulus of 31.71~mag for the Antennae,
as found by Schweizer \etal\ (2008) based on the Type~Ia supernova 2007SR.
This modulus corresponds to a distance of 22~Mpc, hence 1$''$ is equivalent to 107~pc. 

The remainder of this paper is organized as follows: \S2 describes the ALMA observations and reductions; \S3 identifies and labels interesting CO features in the Antennae, tying them back to historical nomenclature;  \S4 examines large-scale structures, highlighting the discovery of a new  ``Long--Thin'' Filament and contrasting it  with SGMC2 and several other regions in the overlap region; \S5 outlines a multi-wavelength, evolutionary sequence designed to provide a framework for understanding the formation and evolution of compact star clusters; \S6 examines small-scale structures in the Antennae with an eye toward evaluating the utility of the framework defined in \S5; \S7 provides a short discussion of age profiles, timescales, and feedback; and \S8 summarizes the results.

\section{Observations}

\subsection{Data and Reductions}

ALMA observed the Antennae as part of its early science campaign (i.e., Cycle~0: project 2011.0.00876). The array was used in its extended
(maximum baseline $\sim$400~m) configuration to  obtain a 13-point mosaic (see Figure~2) positioned to cover the ``overlap region" between the two galaxies. Using the ``Band~7" (345~GHz) receiver, observations of CO~(3-2), 
HCO$^+$~(4-3), HCN~(\mbox{4-3}), and the $\sim$850\,$\mu$m dust continuum emission were made simultaneously.  The entire mosaic was observed by ALMA ten times between April and October 2012, with between 14 and 22 antennae operational.

One of the two quasars J1215-175 or J1159-224 was used as a phase
calibrator on each day. Both are $\sim$3.5$^\circ$ away from the
Antennae. Titan was observed in each track and used to set the overall
flux scale of the observations. The accuracy of ALMA's flux calibration
at 345 GHz is estimated to be $\approx$10\%.\footnote{ALMA Technical
Handbook}

We reduced and calibrated the data using  CASA (Common Astronomy Software Applications)\footnote{http://casa.nrao.edu}  following the standard
procedures for Cycle~0 data reduction. The data were imaged using Briggs weighting with the robustness parameter set to 0.5. After initial imaging, the data were iteratively self-calibrated by averaging the CO line emission across a 500~km~s$^{-1}$ channel into a single plane image (to increase the signal to noise). That image was cleaned and the resulting model was used to carry out an additional calibration using a 10~minute solution interval. The RMS in signal-free regions and the peak intensity remained approximately constant (within $\approx$10\%) before and after the self-calibration. The main effect was to remove visible bowling around the bright filaments.
The final calibrated data were imaged on a $600\times1000$ pixel grid of
$0.08\arcsec$ pixels. The CO (3-2) data cube has 5~km~s$^{-1}$ channels and a synthesized beam $0\farcs56\times0\farcs43$. 

Figure~2 shows the CO (3-2) map overlaid on a $B$, $I$, and Pa$\beta$ color image created from \textit{HST} observations. The solid yellow contours show the CO flux (i.e., moment~0).  The dotted yellow lines show the field of view resulting from the 13-point mosaic. We remind the reader of  vignetting of the sensitivity near the edges of the FOV; the dotted line shows essentially the zero flux sensitivity boundary. The current paper will focus primarily on results from the CO maps while Brogan et~al.\ (2014; Paper~3) will describe the ALMA 850~$\mu$m continuum maps and the dense gas tracers HCN~(4-3) and HCO$^{+}$ (4-3).

\subsection{ALMA Maps of CO (3-2) Emission}

Figure~3 shows the color-coded CO~(3-2)  moment~0 (flux), moment~1
(velocity), and moment~2 (velocity dispersion) maps for the overlap region of the Antennae Galaxies.

The intense regions of CO emission shown by the orange colors in the flux map are coincident with the Super Giant Molecular Clouds (SGMCs)
identified by Wilson et~al.\ (2000). The fainter green and blue regions of emission show an intricate array of knots and filaments. The color-coded
velocity map in the central panel shows a mixture of low velocity emission (1350--1500~km s$^{-1}$: blue through green colors), probably  related to the southern component of the Antennae, NGC~4039, and higher velocity emission (1500--1650~km s$^{-1}$: yellow through purple colors) probably related to the northern component of the Antennae, NGC~4038. This topic will be discussed further in \S4.1. The right panel  shows that regions of very low velocity dispersions ($\approx$10~km~s$^{-1}$ -- shown in black) are often associated with  thin filaments, especially in the northern regions, and the larger dispersion regions (30--80~km~s$^{-1}$: light blue through orange) are roughly coincident with the high-flux SGMCs.

Figure~4 shows the CO velocity channel maps with an interval of 
20~km~s$^{-1}$. Dotted lines define regions denoted as filaments in 
Figure~5. Much of the CO emission appears to be associated with these three filaments. 

\section{How Previously Identified Features Map onto the ALMA CO Observations}  

Figure~5 identifies several  regions within the Antennae that will be discussed
in the present paper.  The optical/near-IR background color image shows
regions of emission, mostly star clusters and individual massive stars forming in the Antennae galaxies, as well as dark regions where dust is silhouetted against the bright disk of the galaxies.  The contours show the CO~(3-2) isophotes.

Four regions in Figure~5 were first identified by Rubin et~al.\ 1970 (i.e.,
Rubin~B, C, D, and E---region C/D is designated for the first time in the present paper since it is a prominent CO feature). An important early result from the \textit{Hubble Space Telescope} was the finding that each of these regions consists of several, or even several dozen, individual young super star clusters (Whitmore et~al.\ 1995, 1999, 2010).  These ``cluster-of-clusters'' are called cluster complexes in what follows. A typical cluster complex has a diameter of a few hundred parsecs. 

The ALMA observations show that there is CO emission associated with  all of the Rubin regions,  although this emission is often not exactly coincident with the brightest optical clusters. Instead, much of the CO emission is found associated with dark clouds in the optical image that are on one side of the cluster complexes (e.g., Rubin~B, C, E), or in the case of Rubin~F (near the top of Figure~2 but out of the field of view in Figure~5) on three sides. 
This is an important result.  A cluster complex  does not exhaust all of the available gas in the first $\approx$ few Myr (i.e., the age estimate for  most of the optical clusters in Rubin~B, C, E, F),  but is still making clusters (proto,
embedded, emerging) at least  5--10~Myr after the earliest clusters in the
region were formed.  This topic will be revisited in \S7.3.

In Figure~5 we also label several Super Giant Molecular Clouds (SGMC1, SGMC2, SGMC3, and SGMC4-5) which were identified by Wilson et~al.\  (2000).  The SGMC regions tend to be younger than the Rubin regions, as evidenced by the larger proportion of proto, embedded, and emerging clusters.  In particular, SGMC4-5 includes the brightest IR source in the entire galaxy (WS80), which was first discovered by Mirabel et~al.\ 1998 (see also Whitmore \& Zhang 2002 for a discussion of WS80---they estimate an age of $\approx$1~Myr for this object).  A few features from other papers are also included in Figure~5 (e.g., W10-3 from Whitmore et~al.\ 2010).

Several kinematic bubbles are also present in the ALMA data. An example, associated with Rubin~E shown in Figure~5, is present in the upper left of the 1580~km~s$^{-1}$ panel in Figure~4.  The feature can also be seen with higher velocity resolution in Figure~6.  The central component is brightest in the 1575~km~s$^{-1}$ panel (i.e., the red point source in the upper left) while parts of the ring can be seen in all six of the panels from 1570 to 
1610~km~s$^{-1}$.  Another prominent bubble is found around WS80 (see Figure~5), and can be seen near the south end of Filament \#3 in the 1600 and 1620~km~s$^{-1}$ panels in Figure~4. These bubbles are briefly discussed in \S7.3.

\section{Large Scale Structure in the Overlap Region of the Antennae---CO (3-2) Morphology}

The high sensitivity and excellent spatial and velocity resolution of the ALMA maps allow us to study the detailed morphologies of the molecular gas
in the Antennae. In this section we focus on {\it large-scale} structure and 
profile five different regions in the overlap region. In \S6 we make a \textit{small-scale}, multi-wavelength comparison between knots and clusters in these same five regions.

\subsection{Discovery of a Long, Thin, Low-Velocity Dispersion CO Filament}

One of the most interesting surprises from the ALMA observations is the discovery of a long, thin, low velocity dispersion filament.  This feature is identified as Filament \#1 in Figures~4 and~5,  and will be referred to as the Long-Thin (or L-T) Filament in what follows.  Figure~7 shows the correspondence between the CO (3-2), radio (3.6~cm), optical/near-IR
continuum and line emission.  The background in the left panel is a \textit{BVI} color image taken with \textit{HST}, while in the central panel the background color image combines JHPa$_{\beta}$ images from \textit{HST}. 
The right panel shows a \textit{BVI}H$_{\alpha}$ image using ACS data (Whitmore et~al.\ 2010) in order to highlight the H$_{\alpha}$ distribution for comparison. The H$_{\alpha}$ image  is not included for the other regions discussed in this paper, but can be seen in Whitmore et~al.\ (2010; Figure~1 or Figure~20).  The small-scale features will be discussed in \S6.2. In this section we focus on the large-scale structures.  

The L-T Filament is approximately 3~kpc long and has a width of $\approx$100~pc (i.e., an aspect ratio of $30/1$). It is composed of roughly ten compact knots. In the current paper we use a preliminary catalog from Leroy et~al.\  (2014) which identifies 8~specific knots with a mean velocity of 
1605~km~s$^{-1}$ and an ensemble velocity dispersion of 
$\approx$10~km~s$^{-1}$ (i.e., 11~km~s$^{-1}$ if all the knots are included or 7~km~s$^{-1}$ if the most southerly knot, which may or may not be part of the structure, is removed).  The cloud velocities are shown in green in 
Figure~7. The internal velocity dispersions of the individual knots are also
very low (i.e., $\approx$10~km~s$^{-1}$), as evidenced by the fact that most of the knots are only seen in the 1600~km~s$^{-1}$ panel of Figure~4. 

Figure~6 shows CO channel maps with 5~km~s$^{-1}$ resolution that highlight the very small velocity ranges for both the filament and the individual knots. The  feature is essentially missing at velocities outside the range between 1590 and 1620~km~s$^{-1}$. We note that at lower flux limits there is also some evidence that the L-T filament extends, at roughly the same velocity,  even further to the north, essentially to the edge of the FOV of the 13 field mosaic (see Figure~3).  This ``extended L-T Filament'' is approximately 5~kpc long.

Filaments of CO gas are commonly seen in the Milky Way, although these are
significantly shorter in extent, typically by a factor of $\approx$10. Examples are the Orion Nebula (Bally et~al.\ 1987), the Pipe Nebula (Lombardi et~al.\ 2006; Bergin and Tafalla 2007), and Nessie (Jackson et~al.\ 2010).   A similar linear feature in the LMC is the N159 + southern molecular ridge, a 2~kpc long feature just to the south of 30~Dor (see Fukui et~al.\ 1999; Yamaguchi et~al.\ 2001, and Indebetouw et~al.\ 2008). Filamentary structures in general are seen on a variety of scales in the universe, ranging from superclusters of galaxies  like the ``Great Wall'' in Perseus (12~Mpc; Ramella et~al.\ 1992), to dust filaments in 223~spiral galaxies (1 to 5~kpc; La~Vigne 
et~al.\ 2006), to subclumps of non-star forming gas in the Pipe Nebula (30~pc; Lombardi et~al.\ 2006).  A relevant question is the degree to which the physics responsible for the different scales of filamentary structure are similar or dissimilar. \emph{The L-T Filament appears to be one of the clearest and longest gas filaments known in the nearby universe.}

Interestingly, there is another CO filament in the Antennae with roughly the same mean velocity and orientation as the L-T Filament, as best shown in the
bottom-left panel of Figure~4 (i.e., at a velocity of 
$\approx$1600~km~s$^{-1}$).  It is not quite as distinct and has a much larger ensemble velocity dispersion than the L-T Filament.  This feature is identified in Figure~3 as Filament \#3, and includes SGMCs~2, 4, 5 (see
Figure~5). The fact that it has the same mean velocity as the L-T Filament, and that both filaments are aligned roughly with the nucleus of NGC~4038
(the northern component of the Antennae), and have roughly the same
velocity as the gas that appears to be associated with the NGC~4038 nucleus ($\approx$1640~km~s$^{-1}$; see Amram et~al.\ 1992), suggests that this material probably originated in that galaxy.

We also note that Filament \#2 appears to be oriented in a direction that is roughly perpendicular to Filaments \#1 and~3, and is aligned with the nucleus of NGC~4039, the southern component of the Antennae galaxies. It has a mean velocity ($\approx$1460~km~s$^{-1}$) similar to the velocity
expected for the eastern component of NGC~4039's edge-on disk (i.e., see Figure \#2 of Amram et~al.\ 1992). Hence it is plausible that the material in Filament \#2 originated from NGC~4039.

Based on SV data, Espada et~al.\ (2012) also discovered a $3.5\times0.2$~kpc
molecular filament with a ``beads-on-a-string'' morphology near the nuclear region of NGC~4039, the southern component of the Antennae. From tip to tip the velocities range from 1600 to 1750~km~s$^{-1}$, unlike the small range for the L-T Filament. However, the dispersion within a cloud, and when comparing with respect to adjacent clouds, is only 
10--20~km~s$^{-1}$. Hence in many ways this feature is quite similar to the L-T filament. The slightly wider width (200~pc) and lower aspect ratio (17-to-1) is probably due to the factor of two poorer spatial resolution available for the SV observations. This feature, as well as the beads-on-a-string morphology in general, will be discussed in more detail in 
\S4.4.

Other optical filaments that have similar morphologies are seen in the western loop region of NGC~4038, as shown in Figure~20 of Whitmore 
et~al.\ 2010 (e.g, region~L---note the star formation along the edges of these filaments), and are discussed in \S9.1 of that paper. In addition,  Whitmore 
et~al.\ (2005), using long-slit spectroscopic observations from \textit{HST}, found low ensemble velocity  dispersions (i.e., in the 5--15~km~s$^{-1}$ range) along several linear features in the Antennae.  It appears that filaments are common in the Antennae galaxies. 

\subsection{The Other Extreme---SGMC2: a Roundish, High-Velocity Dispersion Molecular Cloud}

Much of the CO emission in the Antennae detected in the ALMA data 
has a more spherical morphology than the filaments discussed above.  A number of these regions were defined by Wilson et~al.\ (2000), and are
designated as supergiant molecular clouds (SGMC). These sources are identified in Figure~5. Figure~8 shows the case of SGMC2, one of the rounder clouds.

Using a preliminary catalog from Leroy et~al.\ (2014), we find 12~CO knots in this region, with a range of velocities between 1388 and 1631~km~s$^{-1}$, 
much larger than the velocity range from 1574 to 1610~km~s$^{-1}$ observed for CO knots in the L-T Filament.  This highlights the diversity in the properties of different regions in the Antennae. The flux map 
(moment~0) shown in Figure~8 does not capture the positions of all of the
12~knots very well, due both to crowding and the general complexity of the region.  For example, two of the four regions identified in Figure~8 break up into separate velocity components  in the preliminary Leroy et~al.\ (2014) compilation. Table~1 includes only the four CO regions shown in Figure~8 rather than the 12~knots from the preliminary catalog of Leroy et~al.\ (2014). 

One possible explanation for the larger velocity dispersion of the knots in SGMC2 is that it is near the intersection of Filaments \#2 and \#3.  While it is possible that these two features are merely superposed,  we find that the distribution of the velocities in the 12~knots is relatively uniform over the full range from 1390 to 1630~km~s$^{-1}$, rather than showing a bimodal
distribution around the velocities of the two filaments (i.e., 1460 and 
1600~km~s$^{-1}$). This suggests that the large velocity range is not simply due to the superposition of two low dispersion systems, but may  instead be due to star formation triggered by the collision between the two regions.

SGMC2 is also noteworthy for containing one of the best examples of a 
candidate proto super star cluster in the Antennae (designated as the ``Firecracker'' in Figure~8), as is  discussed in Johnson et~al.\ (2014; 
Paper~4).

\subsection{Three Other Regions---SGMC1, SGMC3 + SGMC4/5, and W10-3}

In this subsection we briefly examine the large-scale structure in three other regions  of the Antennae. The small-scale features of these regions are discussed in \S6.

Figure 9 shows the CO, radio and optical/near-IR observations for the
SGMC1 region (encompassing both Rubin~C [SGMC1-ALMA-2 and~3 in the figure] and Rubin~C/D [SGMC1-ALMA-1]), as shown in Figure~5. The image also extends to Rubin~D in the upper-left part of Figure~9, where only limited CO and radio emission is observed.  The white square shows the location of a maser, discussed by Brogan et~al.\ (2010). Using a
preliminary catalog from Leroy et~al.\ (2014), we find 17~CO knots, with a range of velocities between 1375 and 1519~km~s$^{-1}$.  Hence, this region also has a large ensemble velocity dispersion when compared to the L-T Filament, and a level of complexity similar or greater than SGMC2. Five knots are identified in the moment~0 flux map in Figure~9, several of them with multiple velocity components.

Figure 10 shows the CO, radio and optical/near-IR observations for the
SGMC4/5 region. As discussed in Wilson et~al.\ 2000, these two SGMCs
have similar spatial distributions, but different mean velocities
(1530~km~s$^{-1}$ for SGMC4 and 1580~km~s$^{-1}$ for SGMC5).  The
bottom left region is the well-known WS80 region [SGMC4/5-ALMA-4 and~5], which is the strongest ISO source (Vigroux et~al.\ 1996; Mirabel 
et~al.\ 1998), and strongest radio source (Neff \& Ulvestad 2000) in the
galaxy.  It is also associated with the most massive cluster ($7\times 10^{6}\,M_{\odot}$; Whitmore et~al.\ 2010; see also Whitmore \& Zhang 2002).  The white square shows the location of a maser, discussed by Brogan et~al.\ (2010).  Figure~10 also includes the Rubin~B region (associated with SGMC3) on the right side of the image, with limited CO emission but relatively strong radio emission.  Using a preliminary catalog from Leroy 
et~al.\ (2014), we find 15~CO knots in this field (five of them associated with Rubin~B), with a range of velocities between 1469~km~s$^{-1}$ 
(Region~B) and 1657~km~s$^{-1}$ (a small knot between Rubin~B and WS80). 

Figure 11 shows the CO, radio and optical/near-IR observations for the
W10-3 region, first identified in Whitmore et~al.\ (2010).  Using a
preliminary catalog from Leroy et~al.\ (2014), we find five relatively weak CO knots, with a range of velocities from  1454~km~s$^{-1}$ to 
1515~km s$^{-1}$. This region has an ensemble velocity dispersion~= 
25~km~s$^{-1}$, intermediate between the L-T Filament 
(11~km~s$^{-1}$) and the SGMC regions.  W10-3 is the simplest of the five regions we profile in this paper. For this reason it is particularly useful for developing and illustrating the evolutionary framework discussed in \S5.

We note that Table~1 includes only the 14~CO regions shown in Figures~9, 10, and 11 rather than the 37~knots  listed above from the preliminary catalog of Leroy et~al.\ (2014). 

\subsection{Beads-on-a-String?}

A question implicitly introduced by subsections \S4.1 and~4.2, namely
whether most of the CO emission is associated with filaments or roughly spherical knots, may be largely irrelevant. The answer may be both. On a large scale, it appears that most of the molecular gas is associated with roughly linear filaments.  However, as we transition from the large- to small-scale regimes, we note that the \textit{pieces} of the filaments tend to be more spherical, like the small individual knots in the L-T Filament or W10-3, or the larger cluster complexes in several of the SGMCs (i.e.,  Rubin~B, C, D, WS80, ...). This is the well-known ``beads-on-a-string'' morphology  (e.g., Elmegreen \& Elmegreen 1983).  Hence, while the original large-scale building blocks  may be filamentary in nature, after gravitational instabilities become important the gas may coalesce into more spherical structures, forming the beads within the filaments.

A recent study by Espada et~al.\ (2012) has demonstrated the phenomena for the Antennae, using science verification data to study a molecular gas arm emanating from the nuclear region of NGC~4039, the southern component of the Antennae. They find that the molecular clouds appear to line up like ``beads-on-a-string,'' with almost equidistant separations of about 350~pc. While the current paper does not examine this question in detail, we do note in passing that the separations between knots in the five regions examined in this section appear to be in the range 300  to 600~pc, depending on the minimum mass one uses for the individual beads. 

Hydrodynamic simulations performed by Teyssier et~al.\ (2010) have already produced features that look like beads-on-a-string (see their Figure~3) in a galaxy merger designed to look like the Antennae.  They find that turbulent motions trigger gas fragmentation into massive dense clouds.  Gas velocity dispersions increase from about 10--15~km~s$^{-1}$ by factors of two to five. This is reminiscent of the difference between the L-T Filament (and other linear features reported in Whitmore \& Zhang 2004), and the dispersion found in the SGMC2 ($\approx$80~km~s$^{-1}$) and other supergiant molecular clouds.  Similarly, Renaud et~al.\ (2014), using higher resolution simulations, find that compressive turbulence overcomes the
regulating, stabilizing effect of turbulence, resulting in an excess of dense gas.

\section{An Evolutionary Framework  Connecting CO Knots, Radio Sources and Star Clusters}

In this section we outline an observationally motivated classification scheme for star cluster formation and evolution. This will act as a framework for the 
detailed examination of small-scale, spatial correlations between the CO knots, radio sources, and optical/near-IR clusters that will follow in \S6.

Early versions of parts of this classification scheme have previously been presented in Johnson (2002, 2005) and Whitmore et~al.\ (2011). Johnson (2002, 2005) focused on the earliest stages of proto super star cluster production, using an analogy with ultracompact H{\sc ii} regions (UCHII).  Whitmore et~al.\ (2011) focused on both broad- and narrow-band optical emission from star clusters in the nearby spiral galaxy M83, and showed that morphological categories correlated well with quantitative cluster age estimates.  

Other relevant papers are Kawamura et~al.\ (2009) and Miura et~al.\  (2012). Kawamura et~al.\ (2009; see Yamaguchi et~al.\ 2001 for an earlier version of related work) used NANTEN observations of molecular gas
observations in the Large Magellanic Clouds and developed a model with three evolutionary stages (no star formation, small H{\sc ii} regions, H{\sc ii} regions and young star clusters). Assuming the timescales for each stage are proportional to the number of objects, they find that the lifetime of a GMC is 20--30~Myr.

Miura et~al.\ (2012)  used the ASTE telescope to observe CO~(3-2) for 71 GMCs in M33 with 100~pc resolution, and compared with ages for young stellar groups (YSG) from optical data. They define a four-stage sequence (no star formation, H{\sc ii} regions, H{\sc ii} regions and $<$10~Myr young stellar groups, H{\sc ii} regions and 10--30 Myr young stellar groups). They find that the lifetime of a GMC with mass 
$>$10$^{5}\,M_{\odot}$ is 20--40~Myr.

The primary differences between the current discussion and the Kawamura 
et~al.\ or Miura et~al.\  schemes are the use of ALMA data; the addition of radio observations; and the inclusion of older clusters (i.e., with no CO present) in our treatment.

The sequence proposed in the current paper employs four sets of observations; \textit{CO, radio, optical/near-IR continuum, and   optical/near-IR line emission}. One could easily envision similar classification systems with the inclusion of observations at other wavelengths (e.g., sufficiently high resolution IR observations). The current sequence is designed to include observations with spatial resolution $\approx$0.5$''$ or better, roughly matching the ALMA observations.  In general, each stage of the sequence represents the \textit{addition (or omission) of one or more of the four observational characteristics when attempting to detect the object}, as defined in the matrix shown in Figure~12.  Exceptions are the transitions from stage~0 to~1 (i.e., diffuse to compact CO), and the ambiguity introduced by stochasticity (discussed in Stage~2 below), which leads to the use of the word ``Possible'' for evaluating the presence of line emission in the matrix. 

For closer galaxies, where spatial resolution may be  sufficient to distinguish sub-stages, more detailed versions of the classification system are possible. For example, the ability to resolve a bubble in the optical/near-IR line emission provides the opportunity to  subdivide stage~4 into younger and older sub-stages (i.e., 4a and  4b), as demonstrated by Whitmore et~al.\ (2011). Similarly, if resolved stars can be seen in intermediate-age clusters, this provides a distinction with older clusters which appear fuzzy and diffuse (i.e., Stages~5a and~5b from Whitmore et~al.\ 2011).  A qualitative description of the classification system is outlined below, including a brief
description of the likely physical state of the objects in these observationally determined stages. A quantitative treatment using the small sample in the current paper is presented in \S7.1.

\emph{Stage 0 (diffuse giant molecular clouds)} [turbulent equilibrium] -- Knots of diffuse CO emission are detected, but no radio, or optical/near-IR emission is observed. These regions generally appear as dark regions of dust in 
optical images.

\emph{Stage 1 (protocluster, $<$0.1~Myr)} [high pressure, gravitational collapse] -- Compact CO emission is detected, but no radio or optical/near-IR emission is observed. This stage is expected to be very short lived (i.e., $<$0.1 Myr), hence very rarely observed.

\emph{Stage 2 (embedded cluster, 0.1--1~Myr)} [onset of star formation] -- Thermal radio emission is detected along with CO emission.  No optical/near-IR continuum is observed but weak line emission may be present if an O~star has formed. Low mass clusters may never form an O~star  (i.e., due to stochasticity; e.g., see Fouesneau et~al.\ 2012), hence the use of the word ``Possible''  in this column in Figure~12. The very recently  formed cluster is deeply embedded in its natal gas during this stage.

\emph{Stage 3 (emerging cluster, 1--3~Myr)} [removal of gas and dust] -- The very young cluster is observed primarily in the radio, optical/near-IR emission lines, and faintly in the optical continuum. There may still be weak CO associated with the object, but in many cases the CO is from an adjoining GMC or protocluster.  The cluster is moderately extincted by dust, with $A_V$ typically $\gea$1~mag.

\emph{Stage 4 (young cluster, 3 Myr--10~Myr)} [ISM feedback] -- The cluster is increasingly observable in the optical and in optical/near-IR emission lines (the latter typically in the form of bubbles) due to the removal of much of the natal gas and dust by feedback. CO emission is  gone but weak radio continuum emission is still observed in many cases.  The $A_V$ values have dropped to $\lea$1~mag.  

\emph{Stage 5 (intermediate/old clusters, $>$10~Myr)} [spectral dimming, evaporation] -- The cluster is observed in the optical/near-IR, although it has faded, and ionized gas is no longer observed since the massive stars have evolved into stellar remnants. The cluster slowly loses stars due to 2-body encounters (i.e., evaporation -- also see Fall \& Chandar (2012) for a discussion of various other star cluster destruction processes). 

\section{Small-Scale Structure  in the Overlap Region of the Antennae---A Multi-wavelength Comparison}

With an eye towards evaluating whether the evolutionary classification
system introduced in \S5 can act as a useful framework, we now turn to a more detailed comparison of the locations of molecular gas (traced by CO knots), deeply embedded, recently formed stellar clusters (traced by thermal radio sources), and older clusters (observed in optical/near-IR continuum and line emission).  In order to compare the locations of these sources in the Antennae, we first need to match the \textit{HST} and VLA coordinate systems to the ALMA observations. This was done by matching Pa$\beta$ emission throughout the Antennae system to common features in the 3.6~cm radio emission and \textit{HST} observations. We estimate relative astrometric precision between the three systems is better than 
$\approx$0.1$''$. The VLA data was taken from archival  project AN079, and  has roughly 15 microJy/beam sensitivity.
 
Table 1 provides positions, CO peak and velocities, an estimate of radio peak, and information about clusters in the vicinity of 26~regions in the five areas discussed in the current paper (i.e., as identified in Figures~7 through 11).  
The CO~peak value and velocities were measured with the CPROPS program (Rosolowsky \& Leroy 2006). The radio strength (peak value) estimates were determined from a visual estimate based on the contours shown in Figures~7 through 11 (see Notes to Table~1 for more details). The cluster properties were taken from Whitmore et~al.\ (2010).
Only clusters brighter than 24 in the $I$-band, and within a 24 pixel $\times$ 24 pixel box (i.e., $1.2''\times1.2''$, or 130~pc~ $\times$ 130 pc) centered at the positions listed in Table~1 are included. This box size was chosen in order to roughly match a value of twice the  $0.56''\times0.43''$ synthesized beam size.

The order of discussion in this section has been altered compared to \S4, starting with the clearest example, W10-3 (Figure~11), and finishing with a case which is probably dominated by superpositions, SGMC2 (Figure~8).

\subsection{Multi-wavelength Comparison in W10-3}

We begin our comparison with W10-3, as defined in Whitmore et~al.\ (2010). This presents a relatively simple, isolated region (see Figure~5), and hence is more easily interpreted than the more complex L-T Filament region or the SGMCs.

Figure~11 shows the correspondence between CO knots (green contours), optical star clusters (orange) identified by Whitmore et~al.\ (2010), and radio
continuum emission (3.6~cm) from Brogan et~al.\ (2014) (yellow contours).  

There appear to be clear examples of correspondence between CO emission, radio emission, optical and near-IR continuum, and line-emission for three objects (i.e., W10-3-ALMA-2, 3, and~4).  Using the system defined in \S5, and the matrix shown in Figure~12, these are designated as Stage~3 (emerging clusters) in the right panel of Figure~11,  since they show CO, radio, and optical emission. We note the young ages (1.0~Myr) and moderate extinction (0.9 to 3.3~mag) for the clusters associated with these CO clouds, as determined in Whitmore et~al.\ (2010) based on comparisons  between observed luminosities in different filters and predictions from stellar
evolutionary models.  These age and extinction estimates  are in good agreement with the expectations from the classification scheme.  In particular, we note that the presence of Pa$_{\beta}$ (purple in the right panel of Figure~1) and H$_{\alpha}$ (not shown---see Figure~1 from Whitmore et~al.\ 2010) are consistent with the designation as Stage~3 objects. 

The remaining CO knot (W10-3-ALMA-1) has no optical or radio counterparts, but instead appears to be associated with a dense cloud of dust. We designate this as a Stage~0 (diffuse GMC) object, since it appears
to be a diffuse rather than compact CO cloud. The two other slightly older clusters (6 and 7~Myr) have no CO or radio emission, but do have strong
optical continuum and weak line-emission.  Following the definitions in Figure~12 we designate these two objects as Stage~4 (young clusters).

Hence, the classification scheme provides a good description for W10-3. The only slight issue is that according to the matrix in Figure~12, we might expect faint radio emission from the young clusters (Stage~4), as we will find in many other cases that follow. The lack of radio emission for these young clusters might reflect the fact that radio emission is quite weak for this region in general, or might result from the fact that the clusters tend to be closer to 10~Myr in age than to 3~Myr.  We also note that the cluster masses range from $6\times10^4\,M_{\odot}$ for the cluster associated with W10-3-ALMA-3, to $1.7\times10^5\,M_{\odot}$ for the cluster associated with W10-3-ALMA-4. These are relatively low-mass clusters, which may also explain the lack of radio emission.

We find that Region W10-3 provides a clear and compelling story line of how clusters form and evolve, starting from a dense molecular cloud (Stage~0), including several emerging clusters with CO+radio+optical emission 
(Stage~3), and also including young clusters free of CO and radio emission (Stage~4). We now turn to the more populated regions to see if we can identify other evolutionary stages.

\subsection{Multi-wavelength Comparisons in the L-T Filament and SGMCs 1, 3, 4, and 5}

Figure 7 shows the correspondence between CO knots, optical star clusters, and radio continuum emission  for the L-T Filament. By including a few regions to the side of the L-T Filament (i.e., the string of CO~knots with velocity information running roughly vertically down the middle of the left and central panels), we are able to illustrate all six stages of the evolutionary framework.

Focusing on the central panel, and starting with Stage~0 (diffuse GMCs), we show three regions with only diffuse CO emission; no radio, optical continuum, or line-emission is clearly associated with these regions.

Near the top of the L-T filament we have designated LT-ALMA-1 as a
Stage~1? (protocluster), since we find compact CO, but no radio or optical/near-IR continuum or line emission. The reason for the ``?'' is
twofold: 1)~the radio emission may be too faint to observe (as discussed in the previous section and the notes to Table~1), and 2)~there are some optical clusters just to the east (left) of the CO~knot that may or may not be associated with the CO emission. This spatial offset is a relatively common occurrence, perhaps showing where the center of the CO emission was a few Myrs ago.

A single example of a Stage~2 object (embedded cluster) is shown at the bottom of the L-T Filament (i.e., LT-ALMA-8). Here we find CO and
radio emission, but no optical counterpart, although like LT-ALMA-1,
there are clusters on the edge of the region.

Seven objects designated Stage~3 (emerging clusters) are shown, some including radio emission and some where it is presumably below the detection threshold. Clusters with optical/near-IR continuum or line-emission (see also the H$_{\alpha}$ image in the right panel) are seen in six of the seven 
Stage~3 objects. The only exception is LT-ALMA-2, where cluster candidates are present, but they are fainter than the 24th magnitude threshold.  The values for the cluster ages shown in Figure~7 range from 1 to 3.3~Myr for the Stage~3 objects, and have extinction estimates in the range $A_V = 1.8$ to 4.7, both in agreement with the matrix in Figure~12. We note that the age and extinction values listed in Table~1 are the mean values for the clusters within a $1.2\times1.2''$ box, and hence are generally similar, but slightly different than the single objects shown in the Figures.

Examples of Stage 4 (young cluster; 3--10~Myr) and Stage~5 (intermediate/old cluster; $>$10~Myr) are also shown in Figure~7. These do not appear to be associated with the L-T Filament. Note the radio emission associated with the Stage~4 object in the upper left. This is another case where it is near, but slightly offset from the nearby CO emission. The estimated extinction value for the Stage~5 cluster is $A_V = 0.0$, supporting the interpretation that this is a foreground cluster, almost certainly
unrelated to the much younger L-T Filament.

We now examine SGMC1 (Figure~9) and SGMC3/4/5 (Figure~10).  In
general, the proposed classification system works well for both of these regions, with objects ranging from Stage~0 (diffuse GMCs) to Stage~4 (young clusters) in SGMC1, and from Stage~0 to Stage~5 (intermediate/old clusters) in SGMC3 + SGMC4/5.

In Figure~9, we find that the correspondence between the CO and radio
emission is good in parts of SGMC1 (e.g., the high flux regions in the
southern component), but show interesting differences in other regions. In particular, while there are radio knots on either side of the strong CO emission in Region C/D (one coincident with a maser; Brogan et~al.\ 2010), there is no clear radio emission associated with the CO knot itself. We also find more evidence that radio emission can be associated with slightly older clusters (e.g., a 4~Myr cluster in Rubin~D in the upper left and an 8~Myr cluster on the right in Figure~9).

Similar results are found in the SGMC3 (Rubin~B) and SGMC4/5 regions
shown in Figure~10, with the radio and CO components well aligned in
some regions (e.g., Region~B), and close but not coincident for other
regions (i.e., SGMC4/5-ALMA-4 and 5). We note that some of the most
massive clusters ($>$$10^{6}\,M_{\odot}$) in the galaxy are in Rubin~B and
WS-80 (Whitmore et~al.\ 2010).  As is the case with the L-T Filament,
and SGMC1, the Stage~4 (young clusters) and Stage~5 (intermediate/old
clusters) in the fields do not appear to be associated with the CO emission.

Hence, essentially all of the regions identified in the L-T Filament, and in SGMCs~1, 3, 4 and 5, appear to fit into the proposed classification system quite well. While the relatively high detection threshold in the radio results in some ambiguity for Stage~2, 3 and~4 regions (e.g., in the northern part of the L-T filament), our expectation is that most of these regions will be found to have weak radio emission when deeper radio observations become available. 

\subsection{Multi-wavelength Comparison in SGMC2}

Figure~8 shows the comparison between CO, optical/near-IR, and radio
observations for the SGMC2 region discussed in \S4.2.  

While there appear to be candidate cluster counterparts in two of the four CO
knots in Figure~8 (i.e., SGMC2-ALMA-1 and 3), their physical correspondence is less certain than in the other four regions examined in this paper. For example,  there is little correlation between the CO knots and the Pa$_{\beta}$ emission (purple in Figure~8), unlike what we find in the other regions discussed in \S6.1 and~6.2. It seems likely that some, or possibly all of the apparent correspondence between clusters and CO~knots are due to random superpositions in SGMC2. The crowding and complexity of the region, along with the possibility of extensive extinction, make it difficult to reliably assign stages for this region. 

This is an important point in its own right; that it may not be possible to reliably use the classification system in all parts of the Antennae, or for all galaxies in general. Although we have attempted to label the CO~knots with appropriate stages in Figure~8, we consider them less secure than all four of the other regions we have examined in this paper. A footnote, 
(?)$^{\rm d}$, has been added to  several of the stage estimates for objects in SGMC2 in 
Table~1 for this reason. 

A final note is in order concerning the ``Firecracker'' (SGMC2-ALMA-4; see the left panel of Figure~8).  While there is some diffuse radio emission (i.e., yellow contours) at the location of the CO knot (green contours), it does not appear to be associated with the very compact CO knot itself, but
instead appears to be part of a broad distribution of radio emission associated with the strong radio component to the north. For this reason the designation for this region is Stage~1 (protocluster).  Herrera et~al.\ (2012) find a high H2/CO line ratio in this region, supporting the interpretation as a protocluster. See Johnson et~al.\ (2014) for a more detailed discussion of the Firecracker.

\section{Discussion} 

In this section we briefly discuss quantitative measurements (included in Table~1) for 26~regions examined in the current paper, and outline how the proposed classification system can be used to learn about timescales for the various stages, and the process of feedback. A more detailed treatment will be presented in Whitmore et~al.\ (2015; Paper~5).

\subsection{First Look at CO, Radio and Optical Age Profiles}

Figure 13 shows plots of the CO peak, estimated radio strength, $I$-band magnitudes and extinction ($A_V$) versus evolutionary stage for regions listed in Table~1. 

Before discussing the figures, we note that a relatively large scatter is expected in these plots due to the fact that: 1)~some of the clusters are likely to be superpositions; 2)~closely adjoining regions (i.e., within the 24~pixel 
[1.2$''$]  box) may be in different evolutionary stages, 3)~the limited spatial resolution (especially for the radio emission) may make it difficult to distinguish whether there is an enhancement for regions that are very close to each other (e.g., SGMC3/4/5-ALMA-4 and -5 in Figure~10).

Selection effects will also affect the results. For example, regions with radio flux, but no CO flux, are not included in Table~1 since this is a CO-selected sample. This artificially truncates the tail toward older clusters in the radio strength figure (bottom right panel in Figure~13). In reality, we see several cases of Stage~4 regions that have radio flux but no CO flux (e.g., Rubin~D region in Figure~9). Similarly, Stages~4 and~5 are not represented in the
other panels in Figure~13, not because there are no objects but because we focus on the CO regions in the current paper and only include a few of the Stage~4 and~5 objects for illustrative purposes.  In Paper~5, using a larger sample with the entire overlap region included, we will construct a larger and more uniform sample which mitigate these effects.

We begin our discussion of age profiles in Figure~13 with the plot of CO peak
vs.\ stage shown in the bottom left panel. As predicted by the matrix in Figure~12, we find low values of CO peak at the youngest age (i.e., Stage 0 -- diffuse GMCs), before star formation has begun. Most of the strongest CO sources are found in Stages~1 and~2.  Presumably, denser regions of gas lead to stronger CO peak, and more intense star formation, which then removes the gas at later stages.  By the time Stage~3 (emerging cluster) is reached, the CO peak is again quite weak, either because the gas has been converted to stars or has been expelled by feedback. As expected, Rubin~C, WS-80 and the ``Firecracker'' are three of the brightest CO~regions, as labeled in Figure~13.

The radio strength (bottom-right panel in Figure~13) shows a similar age profile, with a peak for Stage~1 and 2 regions.  The onset of star formation at these stages results in the generation of free-free radio emission. However, it should be kept in mind that the value for Stage~0 is zero by definition (see Figure~12), making it hard to interpret the age profile for young ages. No value for the mean is shown for this reason.  Another concern is that due to selection effects, as discussed above, the tail toward later stages is truncated. In reality, we find several stage~4 objects with relatively strong radio emission (e.g., Rubin~D in Figure~9), as implied by the matrix in Figure~12. 

The $I$-band flux (upper-left panel in Figure~13) increases from essentially zero for Stage~0 (the single region that shows a value is likely to be a superposition---the other three values have been set to 25th magnitude to make them visible in the figure), to the three brightest (apparent) $I$-band magnitude regions in Stage~3, as listed in Table~1. When corrected for extinction, as shown in Figure~13, the age profile from Stage~1 to~3 is flatter, with WS-80 and Rubin~C as the two brightest regions. After gas expulsion removes the dust for Stage~4 and~5 objects, the decline in optical
magnitudes is driven by stellar dimming, as well documented in a large number of studies (e.g., Fall et~al.\ 2005). The selection effect discussed in the previous two paragraphs keeps us from demonstrating this in the current paper. We also note that the faintest optical clusters in this diagram tend to be
objects we suspect are superpositions, since they are drawn from the full sample
of clusters, most of which are faint. 

Extinction values (upper-right panel in Figure~13) drop from Stage~1 to~3 as
expected, and as implied by the stage names (protoclusters $>$ embedded clusters $>$ emerging clusters). The mean value of $A_V$ for stage~3 is 2.6 for clusters with a mean age of 2~Myr (after eliminating the value of 
130~Myr which is probably from a superposed cluster). This is in good agreement with estimates from Whitmore \& Zhang (2002) and Mengel 
et~al.\ (2005). The two Stage~1 regions in WS-80, with 
$A_V\approx6$~mag, have the highest values of extinction, as expected.  

To summarize this section, as with the qualitative treatment in \S6, a brief quantitative assessment of the specific objects discussed in the current paper provides evidence for the age profile trends predicted by the matrix in 
Figure~12 and the discussion in \S5. A more extensive treatment will be included in Paper~5.

\subsection{A Quick Census and Comparison with Predicted Fractions Based on Timescales}

The estimated timescales for the different stages outlined in \S5 are
dramatically different, ranging from 0.1~Myr for Stage~1 (protoclusters---where the relevant timeframe is the free-fall time, see Johnson et~al.\ 2014 for a calculation based on the Firecracker) to billions of years for Stage~5 (intermediate/old clusters---where the relevant timeframe is the evaporation time---e.g., see Fall and Chandar 2012).

In principle, using the framework developed above, and assuming a roughly constant cluster formation rate, we can determine whether our expected time scales are reasonable by simply counting the number of objects in each of the six stages. In practice, many complications arise. These include selection and completeness effects; the possibility that many of the clusters may be destroyed with time (e.g., see Whitmore et~al.\  2007); the fact that the star formation rate is not constant; and low-number statistics.  Nonetheless, if we cautiously forge ahead we can at least make order-of-magnitude estimates, as has been done by Kawamura et~al.\ (2009) and Miura et~al.\ (2012).

Using timescale estimates from the matrix in Figure~12 and discussion in 
\S5, and a sample of 22~regions from Table~1 (i.e., removing stages~0, 4, 
and~5 since these have only been sporadically included in the present paper for illustrative purposes) we predict stages 1/ 2/ 3 should be populated by the
following numbers: [timescale/sample size = 0.1/22, 0.9/22, 2/22] = [3\%, 
30\%, 67\%]. Using values from Table~1, and interpreting values of ``1?'' as~1 (i.e., true protoclusters), we find ``observed'' values = [N/sample size~=  5/22, 5/22, 12/22]~= [23\%, 23\%, 55\%].  If we convert value of 1? to~2 (i.e., assuming that deeper or higher resolution radio observations will eventually detect radio emission from these objects and move them into  Stage~2 -- embedded cluster) the numbers become [N/sample size~= 1/22, 9/22, 
12/22]~= [5\%, 41\%, 55\%]. These latter values would be reasonably close to the predicted values. Clearly, part of the focus for the future will be to
scrutinize the ``1?'' objects from Table~1 more closely to determine if they are
viable Stage~1 (protocluster) candidates.

We caution the reader from taking these numbers too seriously; they are primarily meant as illustrative examples, and to outline the approach that will be used for Paper~5. 

\subsection{Estimating the Age Spread for Star Formation in Cluster Complexes (i.e., the GMC Destruction Timescale}

In this section we use the new classification system, and the SED ages from Whitmore et~al.\ (2010), to estimate the age spread of star formation in cluster complexes. This  also provides an estimate of how quickly feedback from the newly formed star clusters is able to clear away the remnants of the GMC, resulting in the cessation of star formation in the region.  In practice, this  requires us to determine how wide a range of ages (i.e., stages) can be found coexisting in the same cluster complex.

W10-3 (Figure~9) provides an example of a cluster complex with a mix of recent star formation (i.e., Stage~4 clusters with age estimates of~6 and 
7~Myr), ongoing star formation (i.e., Stage~3 emerging clusters with ages around 1~Myr), and possible future star formation (i.e., Stage~0 diffuse GMCs). Hence there is a spread from Stage~0 to Stage~4, equivalent to~6 or 7~Myr in this case.  The other regions studied in this paper all show similar
distributions, with regions in Stages~0 through 4 seen in close proximity, but no cases where a Stage~5 (intermediate/old cluster with age $>$10~Myr) is clearly associated with the cluster complex. Hence, $\approx$10~Myr can be considered a rough estimate for the amount of time it takes to destroy the GMCs in a cluster complex.

Perhaps the only questionable case is near the southern end of the L-T Filament (Figure~7), with a 200~Myr cluster near the CO filament. However, the estimated extinction for this cluster is $A_V =0.0$ (the only example of this in the entire paper) indicating that the cluster  is very likely to be in the foreground, and hence is almost certainly not associated with the CO emission. 

Another interesting case is a $\approx$50~Myr cluster to the west (right) of
SGMC3 (Rubin~B) in Figure~10. Located at a projected distance of about 200~pc from SGMC3-ALMA-3, Whitmore et~al.\ (2010) believe that this cluster may have triggered the formation of many of the clusters in Rubin~B,
based on the presence of a large ring of H$_{\alpha}$ centered on the cluster, with the eastern edge of the bubble  aligned with the region with the youngest clusters  (see Figure~22 and the discussion in \S9.2 in Whitmore 
et~al.\ 2010).  They also discuss the possibility that the cluster is somewhat younger, with an age in the 10--20~Myr range. In any case, this cluster, and the subsequent generations of clusters it has spawned,  may have cleared most of the CO gas out of its local environment out to a distance of 
$\approx$200~pc.

We can broaden the sample slightly by including several regions identified in Whitmore et~al.\ (2010; Figure~17 and Table~9).  In particular, Region~4 (mean age~= 80~Myr), Region~5 (mean age~= 8~Myr), Region~6 (mean age~= 30~Myr), and Region~7 (mean age~= 50~Myr) are largely devoid of CO emission. These regions can be seen as the bright optical clusters in the upper left portion of Figure~2 in the current paper. The only region with a small amount of CO emission is Region~5 (i.e., the CO knot farthest to the upper left in Figure~2). This is to be expected since Region~5 is the only region with an age less than 10~Myr in this sample of four.

Hence, our current estimate for an upper limit for the age spread in typical cluster complexes is $\approx$10~Myr. On that timescale it appears that the clusters have had enough time to expel the natal gas in the local GMC they were born in, or alternatively, the gas has all been converted into stars. This estimate will be refined in Whitmore et~al.\ (2015; paper~5) using a larger sample.   

It is interesting to note that Kawamura et~al.\ (2009) estimate a 20--30~Myr lifetime for GMCs in regions with recent star formation for their LMC data set, and Miura et~al.\ (2012) find the range to be 20--40~Myr in M33, both slightly larger than our estimate of $\approx$10~Myr.

The removal of gas during the formation of super-bubbles around cluster complexes is a likely mechanism for the destruction of GMCs in these regions. Typical bubble velocities in the Antennae are 20--30~km~s$^{-1}$, based on the H$_{\alpha}$ observations reported in Amram et~al.\ 1992, our own ALMA observations in this paper (\S2.2), and H$_{\alpha}$ observations by Whitmore et~al.\ (1995) and Whitmore \& Zhang (2002). Assuming a typical cluster complex radius of 200~pc results in an estimated  crossing time of $\approx$10~Myr. Hence the derived $\approx$10~Myr
estimate for the upper limit of the age spread is roughly consistent with the
hypothesis that ISM feedback from super-bubbles is responsible for the removal of the GMCs, and hence the cessation of star formation in cluster complexes.
		  
\section{Summary and Conclusions}

In this paper we present early (Cycle~0) results based on observations of the Antennae galaxies (NGC~4038/39) using the Atacama Large
Millimeter/sub-millimeter Array (ALMA) to provide a significant improvement in spatial resolution (\textit{fwhm}~= 0.5$''$). Our goals are to present the data; examine some aspects of the large scale structure (e.g, highlighting the discovery of the Long-Thin Filament); examine the small-scale correspondence between CO, optical, and radio (3.6~cm) emission in a few instructive regions; outline a classification system designed to act as a framework for studying the sequence of star-cluster formation and evolution, and briefly examine observed versus predicted timescales.
More detailed quantitative treatments will follow in separate papers. The primary new results are listed below.

1.  We identify a long (3--5~kpc), thin (aspect ratio 30-to-1) filament of CO gas (aka the Long-Thin or L-T Filament), which breaks up into roughly ten individual knots, and has a very low velocity dispersion 
$\approx$10~km~s$^{-1}$ .  Several other filaments, including a recently discovered ``molecular arm'' by Espada et~al.\ (2012), are also discussed.  At the other extreme, SGMC2 consists of an approximately spherical assembly of knots with a much higher velocity dispersion 
($\approx$80~km~s$^{-1}$). Most regions (e.g., the L-T Filament, SGMC1,
SGMC4/5, W10-3) have properties of both, with a ``beads-on-a-string''
morphology.

2. A comparison between the locations of CO~knots, radio emission, and optical/near-IR continuum and line emission in several regions (especially W10-3, a relatively isolated filamentary cluster complex), presents a clear evolutionary story line, with CO-only regions aligned with dust clouds; joint CO+radio regions  associated with very young embedded clusters with ages 
$\approx$1--3~Myr and values of $A_V$  typically  greater than 1;  and clusters with ages $>$3~Myr having no CO~emission.

3. This multi-wavelength comparison leads us to develop an evolutionary framework connecting diffuse GMCs (Stage~0), protoclusters (Stage~1), embedded clusters (Stage~2), emerging clusters (Stage~3), young clusters
(Stage~4), and intermediate/old clusters (Stage~5). This classification
system will be described in more detail in Whitmore et~al.\ (2015; Paper~5), and will be used to classify features throughout the overlap region in the Antennae.

4. A rough check of the timescales for the different evolutionary stages has been made by assuming a roughly constant cluster formation rate and counting the number of regions in each stage of evolution. A reasonable agreement is found, with the main uncertainty being a set of four regions designated as Stage~1?,  with the question mark reflecting the uncertainty in the radio properties of these candidate protoclusters.

5. Using this framework we conclude that the maximum age spread for a
cluster complex (or equivalently, the lifetime of a GMC in a star-formation region) is $\approx$10~Myr, slightly lower than estimates by Kawamura 
et~al.\ (2009; 20--30~Myr) and Miura et~al.\ (2012; 20--40~Myr). This estimate is based on the fact that there are many regions with a simultaneous mixture of objects in Stages~1 (protoclusters) through~4
(young clusters), but no regions that we are currently aware of that have objects in Stages~1 through~5 (intermediate/old clusters).  A likely explanation is that feedback from the recently formed star clusters is able to clear away the remnants of the GMC within a radius of about 200~pc in this period of time.

\acknowledgments
We thank the referee for many useful comments that improved the paper. 
This paper makes use of the following ALMA data:
ADS/JAO.ALMA\#2011.0.00876.S.  ALMA is a partnership of ESO
(representing its member states), NSF (USA) and NINS (Japan), together
with NRC (Canada) and NSC and ASIAA (Taiwan), in cooperation with the
Republic of Chile. The Joint ALMA Observatory is operated by ESO,
AUI/NRAO and NAOJ. The National Radio Astronomy Observatory is a
facility of the National Science Foundation operated under cooperative
agreement by Associated universities, Inc. This paper also used archival 
VLA data from project AN079. This paper is also based on observations taken with the NASA/ESA \textit{Hubble Space Telescope} obtained at the Space Telescope Science Institute, which is operated by AURA, Inc., under NASA contract NAS5-26555.  This research has made use of the NASA/IPAC Extragalactic Database (NED), which is operated by the Jet Propulsion Laboratory, California Institute of Technology, under contract with NASA. This work was supported in part by the National Science Foundation under Grant No. PHYS-1066293 and the hospitality of the Aspen Center for Physics.

{\it Facilities:} \facility{HST, ALMA, VLA}.

\clearpage

\begin{figure}
\begin{center}
\includegraphics[width = 4.3in, angle= 0]{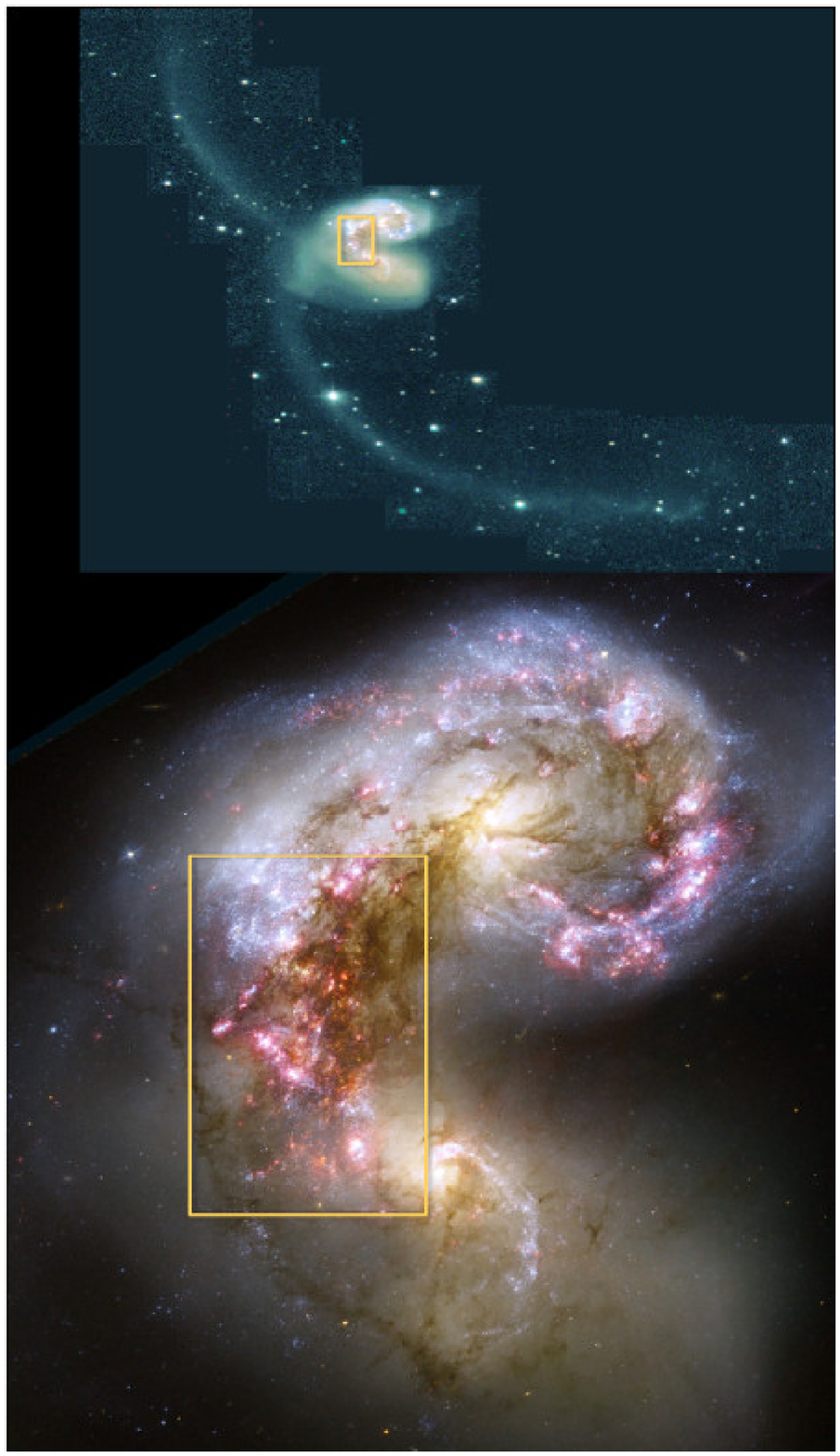}
\end{center}
\caption{Two views of the merging Antennae galaxies.  The top   figure shows a ground-based \textit{BVI} color image, which highlights the long tidal tails (courtesy of John Hibbard).  An optical color image of the main body of the Antennae taken with \textit{HST} is shown in the bottom image (Whitmore et~al.\ 2010).    The yellow box shows the approximate coverage of
the Cycle~0 ALMA observations.}
\end{figure}

\begin{figure}
\begin{center}
\plotone{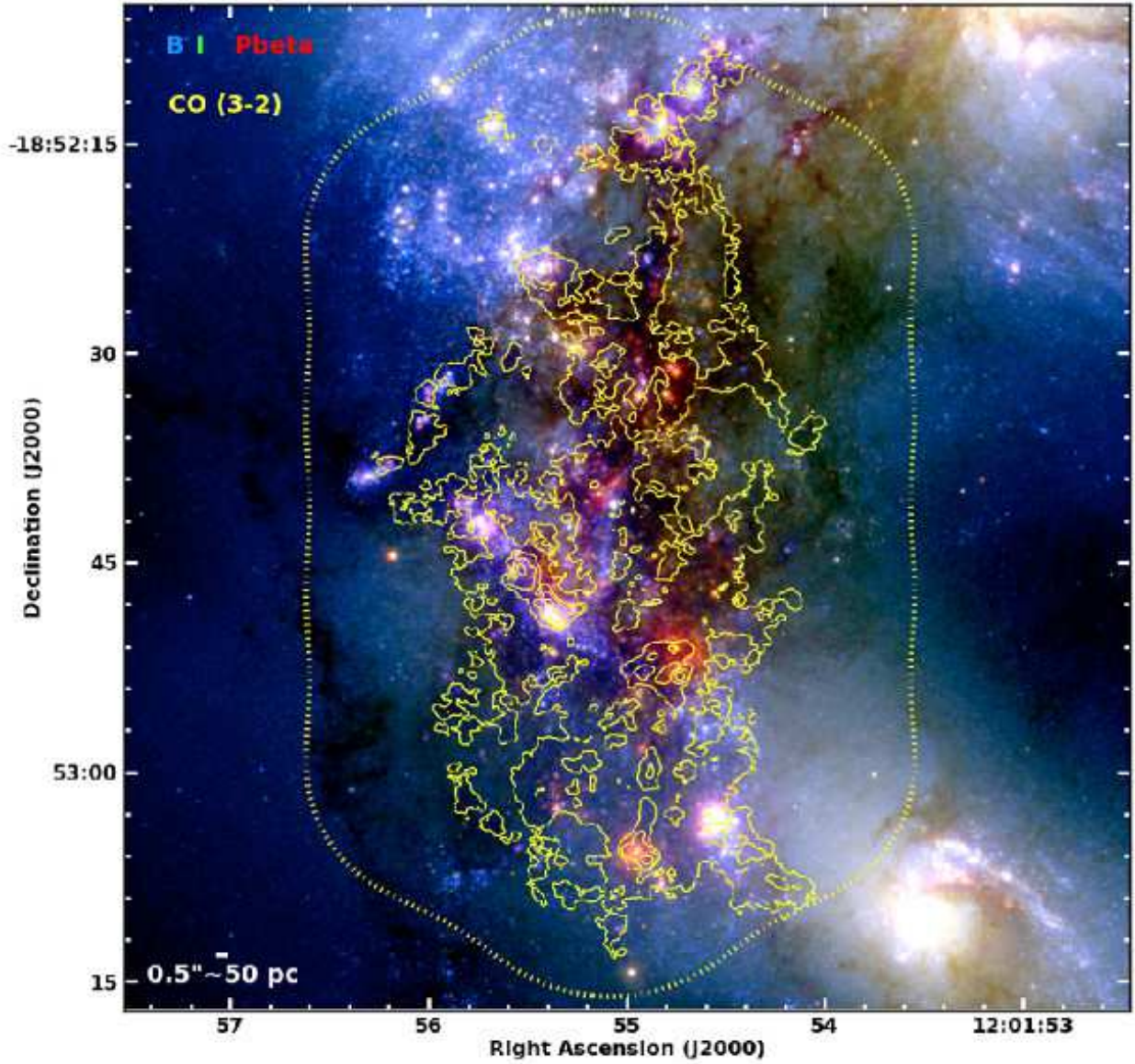}
\end{center}
\caption{CO (3-2) emission flux (moment~0) map superposed on a 
\textit{HST} color image with blue = B, green = I, and red = Pa$_{\beta}$. The contour levels are [4, 200, 400, 800]~K~km~s$^{-1}$. We note that 
0.1~Jy beam$^{-1} = 4.3$~K at the resolution and frequency of these images
(i.e., $0.56''\times 0.43''$).
}  
\end{figure}

\begin{landscape}
\begin{figure}
\begin{center}
\includegraphics[width = 4.4in, angle= -90]{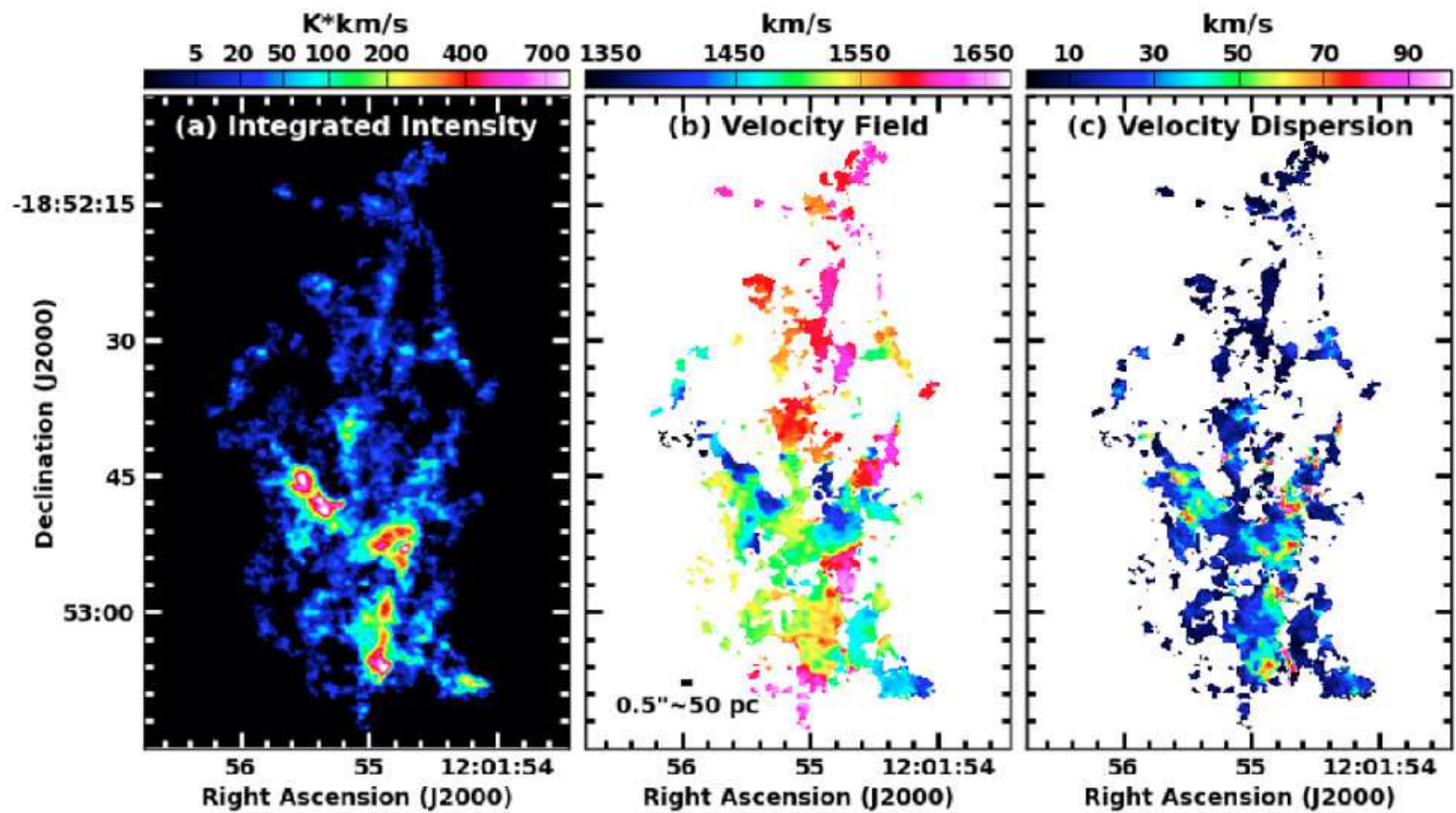}  
\end{center}
\caption{Left: CO (3-2)  flux (moment~0), middle: CO velocities 
(moment~1), right: CO velocity dispersions (moment~2) maps for the overlap region of the Antennae Galaxies. The scale for the color coding is shown along the top. }  
\end{figure}
\end{landscape}

\begin{figure}
\begin{center}
\includegraphics[width = 5.5in, angle= 0]{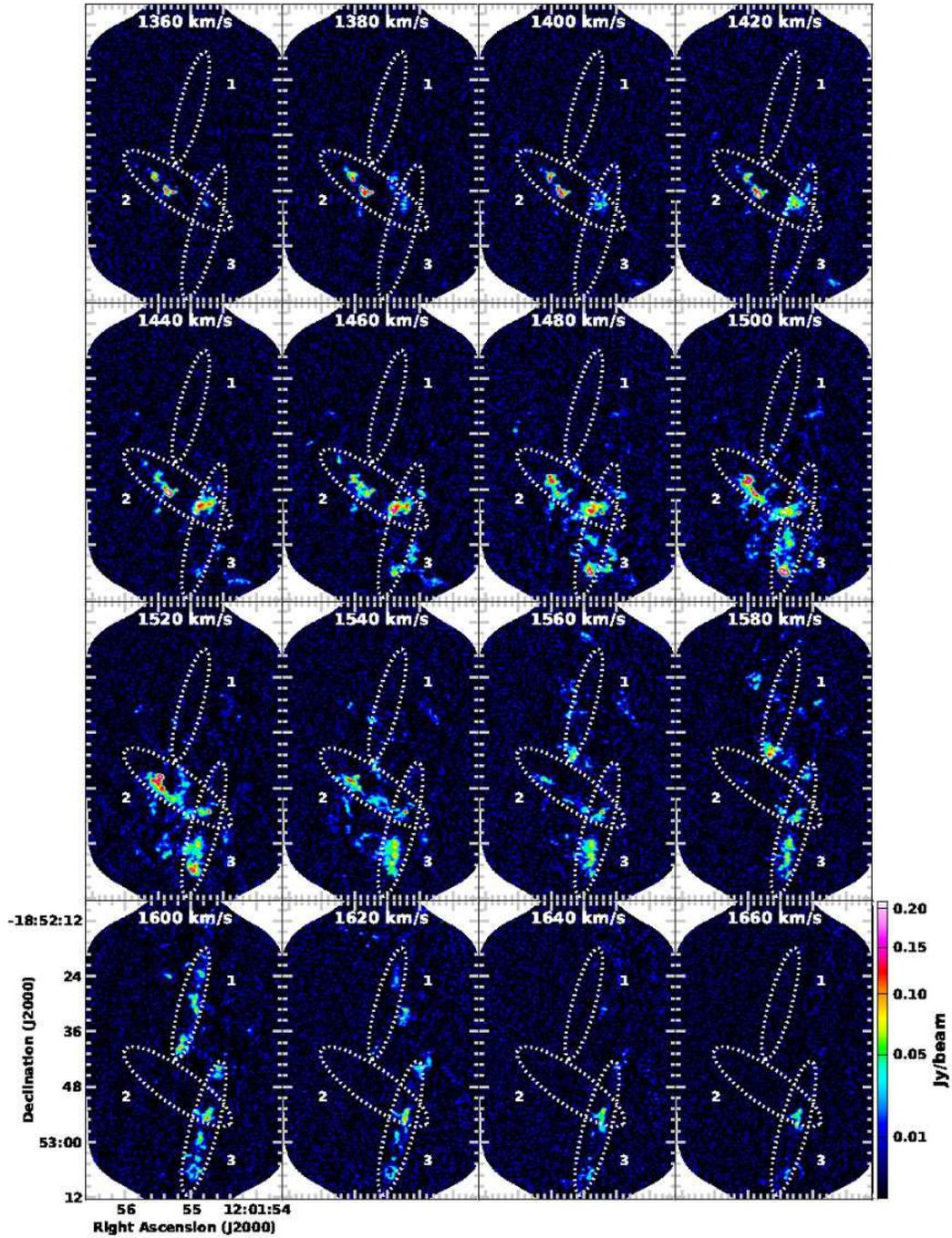}
\end{center}
\caption{Velocity channel maps of the overlap region in the Antennae Galaxies. The locations of Filaments \#1, 2, and~3 (see Figure~5) are  indicated by the dotted lines. The Right Ascension and Declination scales are shown along the bottom left while the scale for the color coding is shown along the bottom right.}  
\end{figure}

\begin{figure}
\begin{center}
\plotone{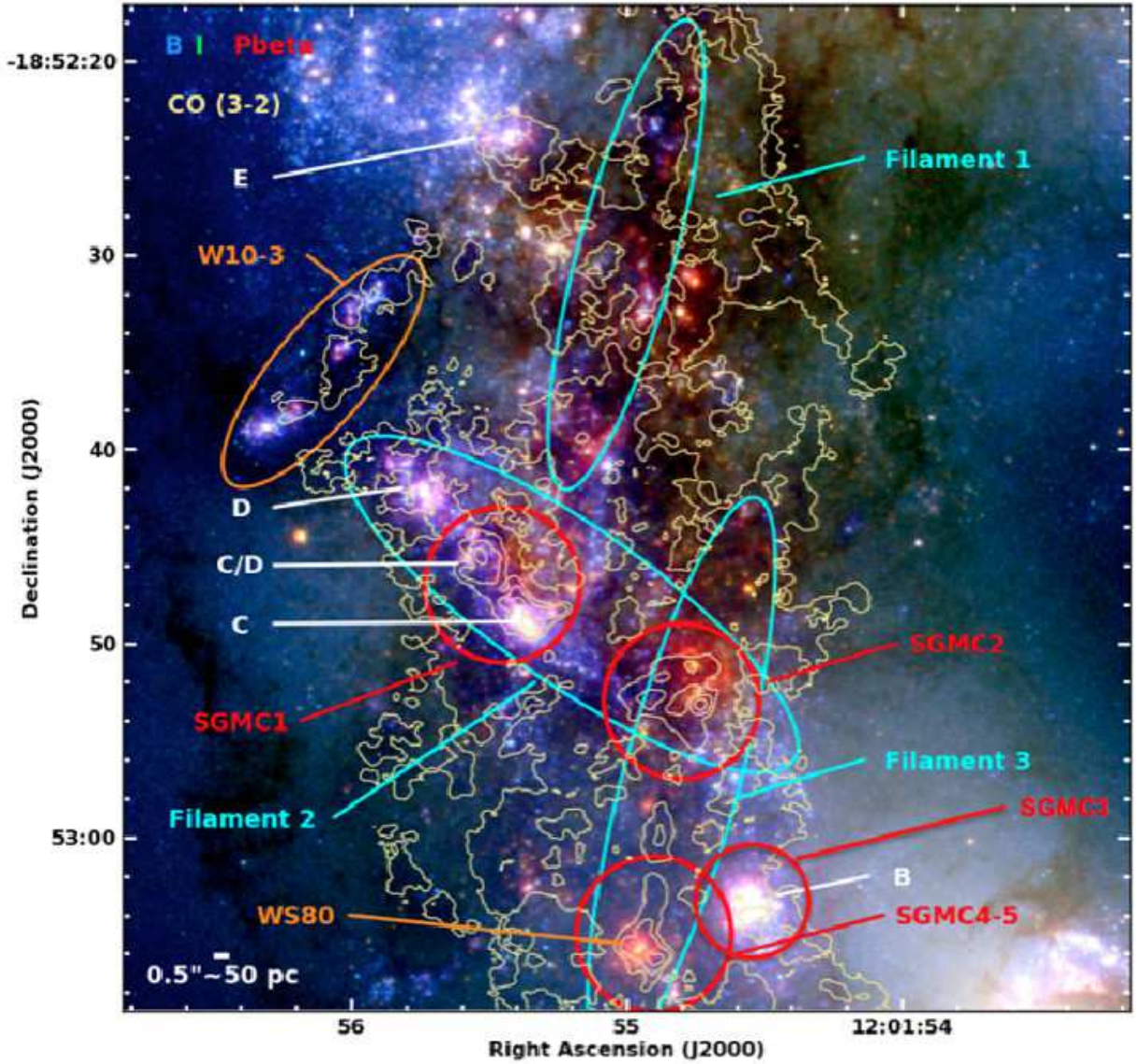}
\end{center}
\caption{CO flux  map (yellow contours) superposed on a \textit{HST} color image with blue = B, green = I, and red = Pa$_{\beta}$. The contour levels are [4, 200, 400, 800]~K~km~s$^{-1}$. Several previously identified features are shown including Super Giant Molecular Clouds (SGMCs; Wilson 
et~al.\ 2000), bright cluster complexes (i.e., B, C/D, C, D, E; Rubin et~al.\ 1970), and two regions of special interest (i.e., W10-3 -- Whitmore et~al.\  2010; and WS80 -- Whitmore and Zhang 2002). Three filaments discussed in the text are also shown.}  
\end{figure}

\begin{figure}
\begin{center}
\includegraphics[width = 6.5in, angle= 0]{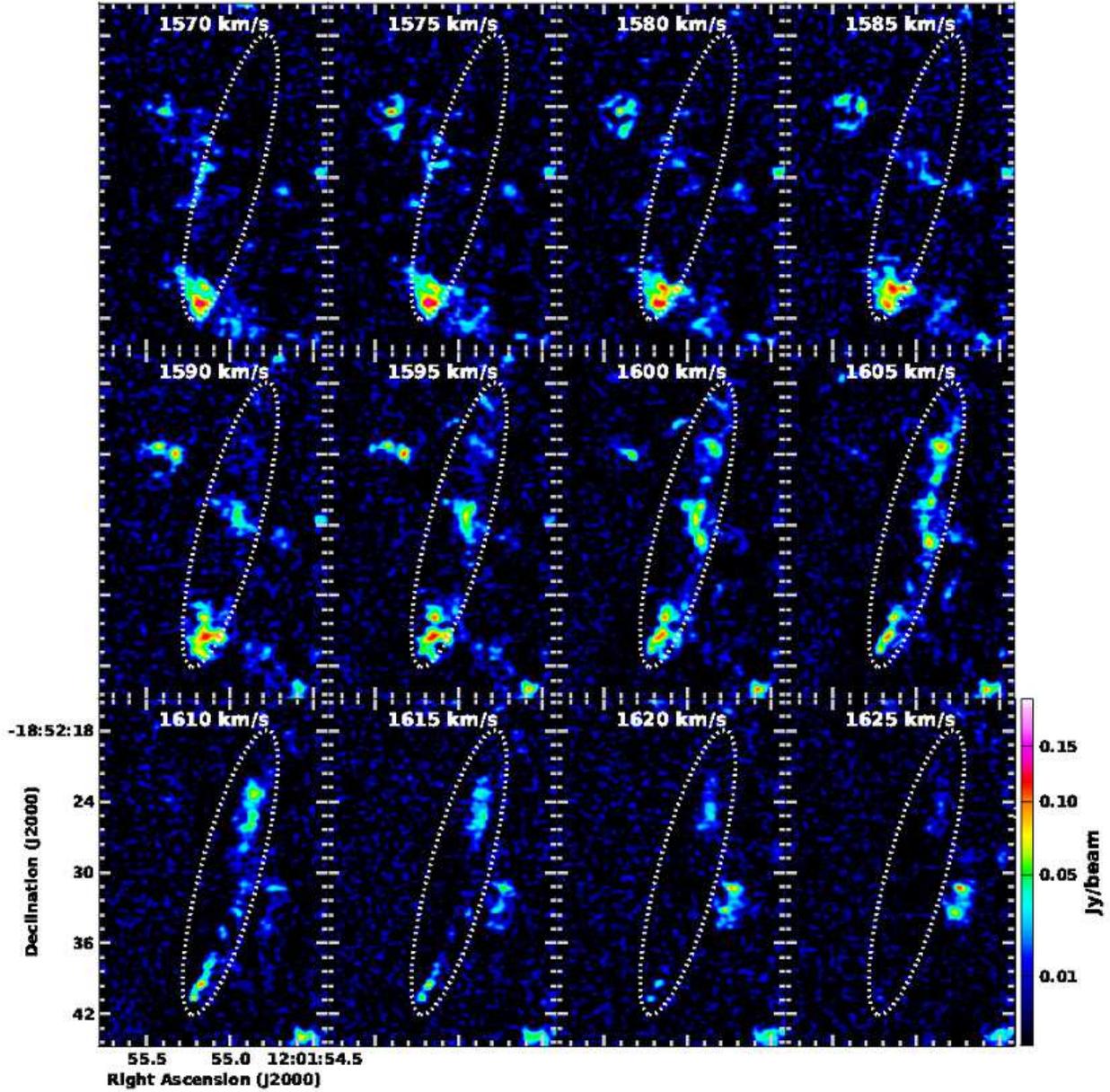}  
\end{center}
\caption{Velocity channel map showing velocities  around the central velocity of the Long-Thin Filament (aka Filament \#1).  The location of the filament
is indicated by the dotted lines. The Right Ascension and Declination scales are shown along the bottom left while the scale for the color coding is shown along the bottom right.}
\end{figure}

\begin{landscape}
\begin{figure}
\begin{center}
\includegraphics[width = 4.2in, angle= -90]{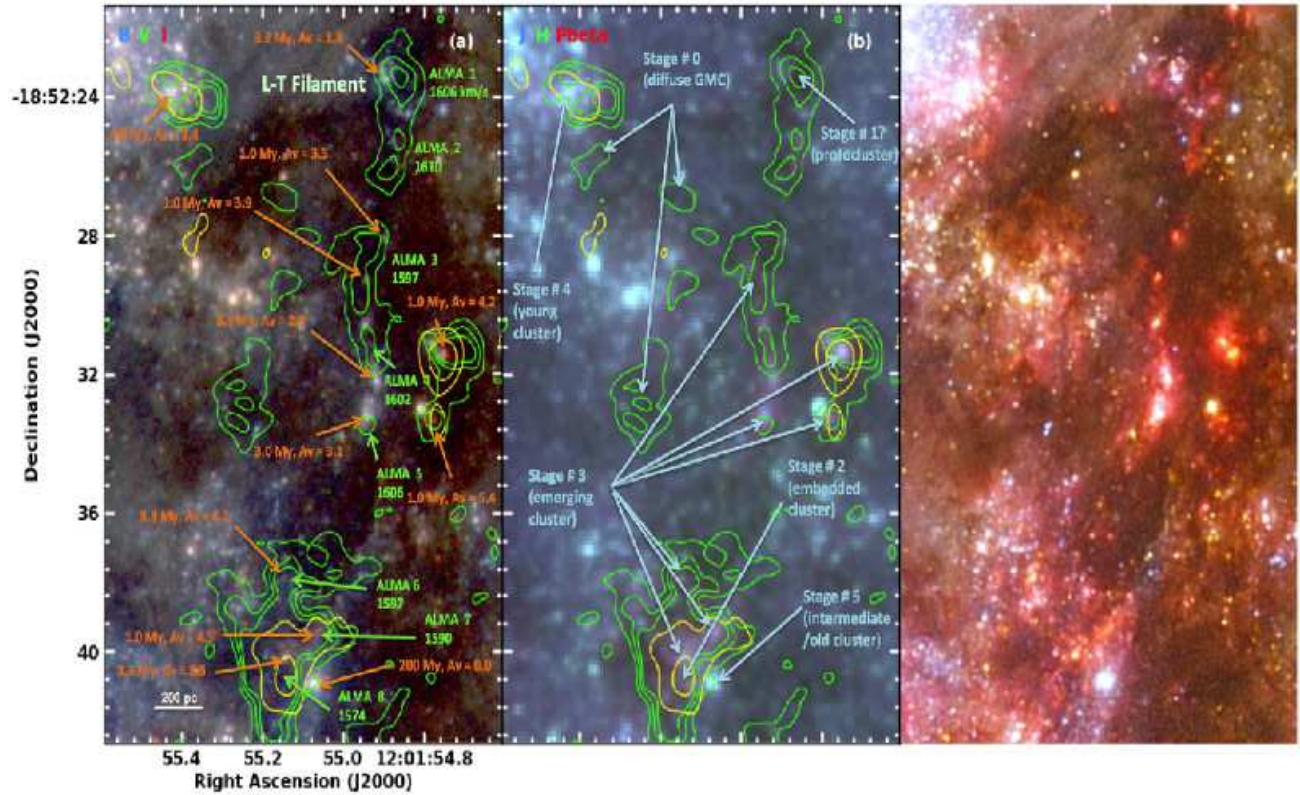}  
\end{center}
\caption{Correspondence between CO (3-2) knots, 3.6~cm radio  emission and optical star clusters in the L-T Filament. The green contours (20, 40, 
60~K~km~s$^{-1}$) show CO emission; yellow contours (50, 150~$\mu$Jy beam$^{-1}$)  show radio emission; and orange provides information about the optical clusters. We note that 0.1~Jy beam$^{-1} = 4.3$~K at the resolution and frequency of these images (i.e., $0.56''\times0.43''$). A 200~pc (1.9$''$) bar is provided in the bottom left to indicate the scale. The mean radial velocities for the CO knots are   listed below the knot ID.  The left image is a composite using  the $B$, $V$, and $I$ \textit{HST} WFC3 images; the center panel uses the J, H and Pa$_{\beta}$ WFC3 images; and the right panel uses the $B$, $V$, $I$, and H$_{\alpha}$ \textit{HST} ACS images.  
See text for more details.
}
\end{figure}
\end{landscape}

\begin{landscape}
\begin{figure}
\begin{center}  
\includegraphics[width = 4.5in, angle= -90]{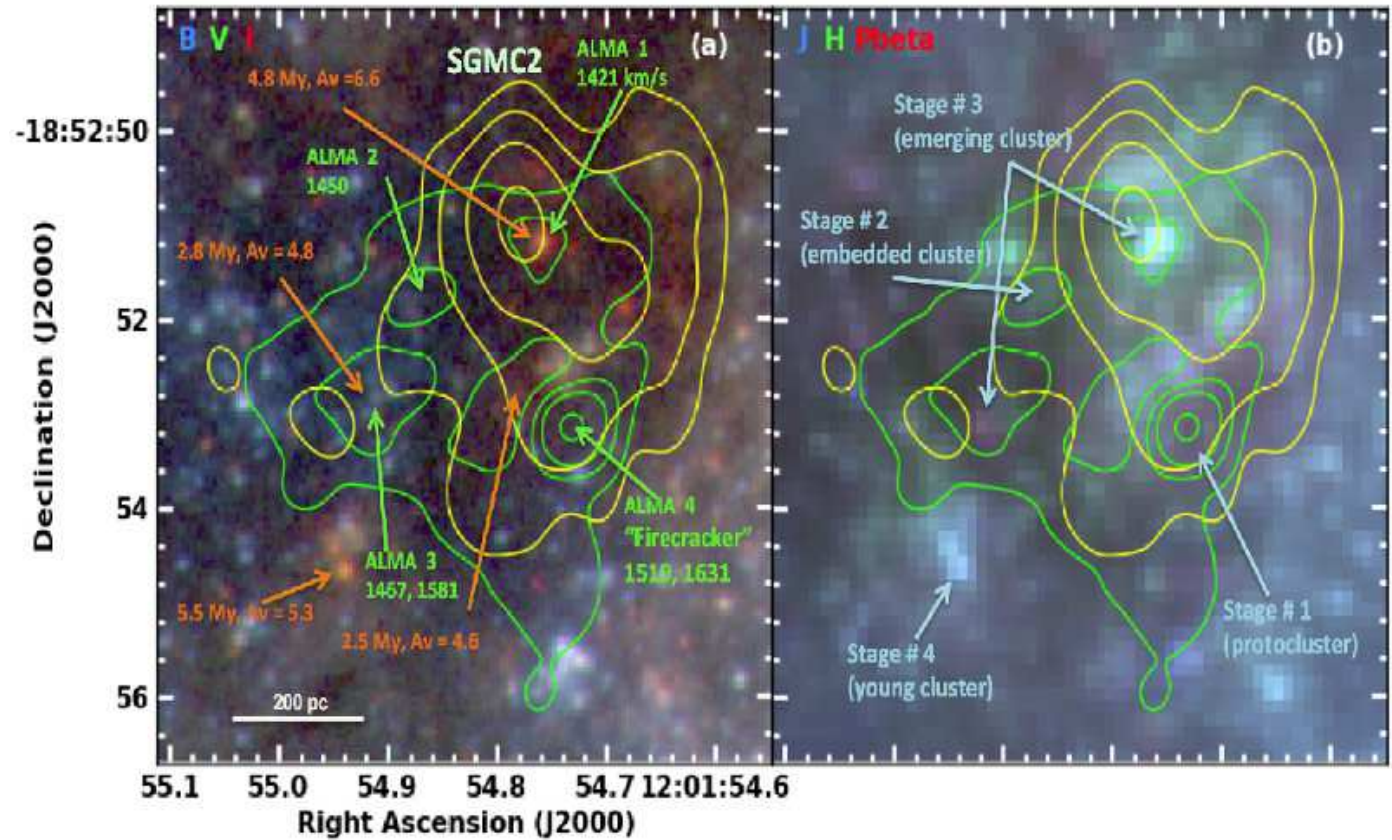}  
\end{center}
\caption{Correspondence between CO knots (green), 3.6~cm radio  emission (yellow) and optical star clusters (orange) in SGMC2.  The green contours (180, 380, 580, 980~K~km~s$^{-1}$) show CO emission; yellow contours (80, 160, 320, 640~$\mu$Jy beam$^{-1}$)  show radio emission. See 
Figure~7 for a more detailed description of the figure. }
\end{figure}
\end{landscape}

\begin{landscape}
\begin{figure}
\begin{center}
\includegraphics[width = 4.5in, angle= -90]{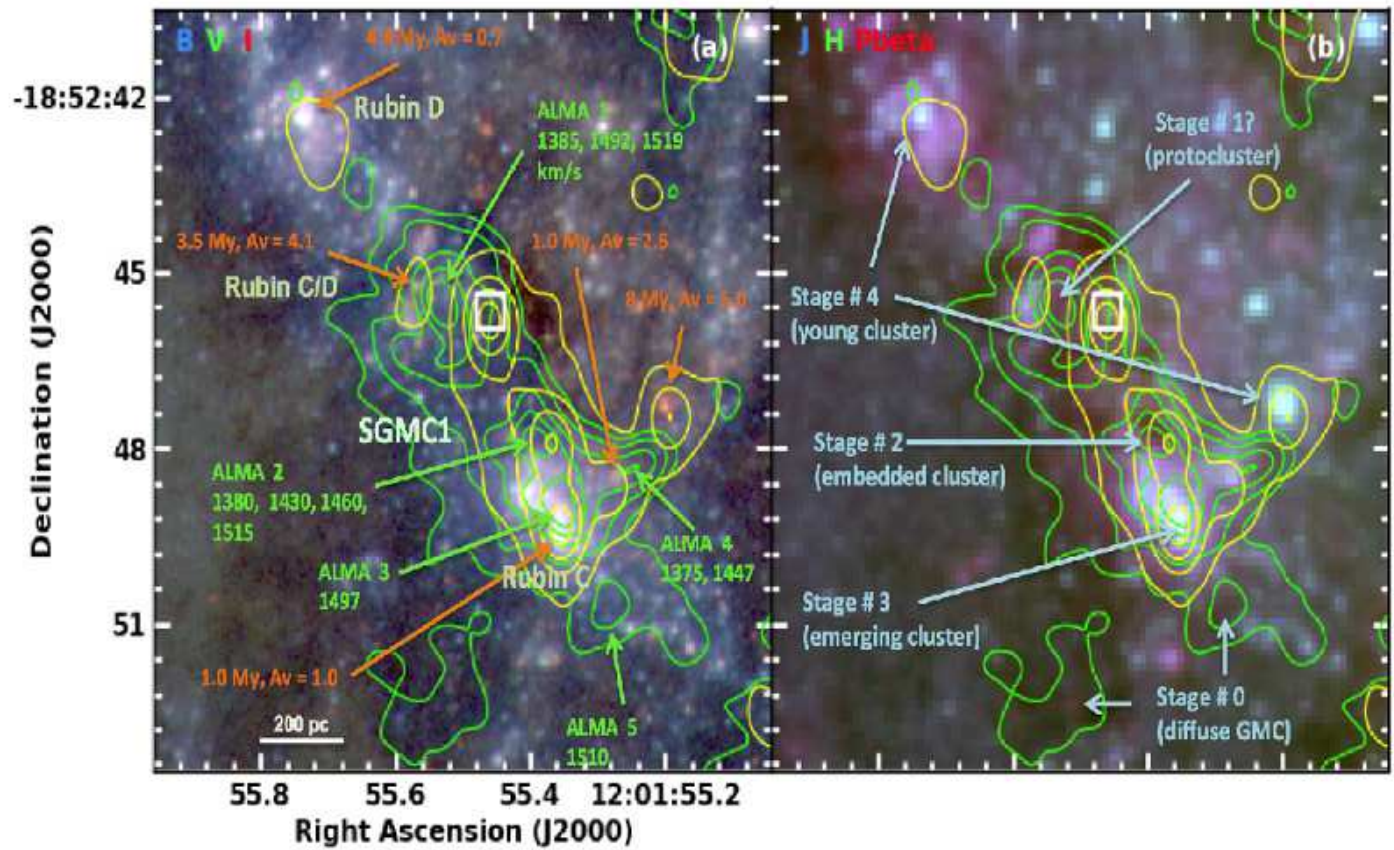}  
\end{center}
\caption{Correspondence between CO knots (green), 3.6~cm radio emission (yellow)  and optical star clusters (orange)  in SGMC1 (including Rubin~C, C/D, and~D). The green contours (70, 170, 370, 570, 770~ K~* 
km~s$^{-1}$) show CO emission; yellow contours (60, 200, 340, 
480~$\mu$Jy beam$^{-1}$)  show radio emission. See Figure~7 for a more detailed description of the figure. 
}
\end{figure}
\end{landscape}

\begin{landscape}
\begin{figure}
\begin{center}
\includegraphics[width = 4.5in, angle= -90]{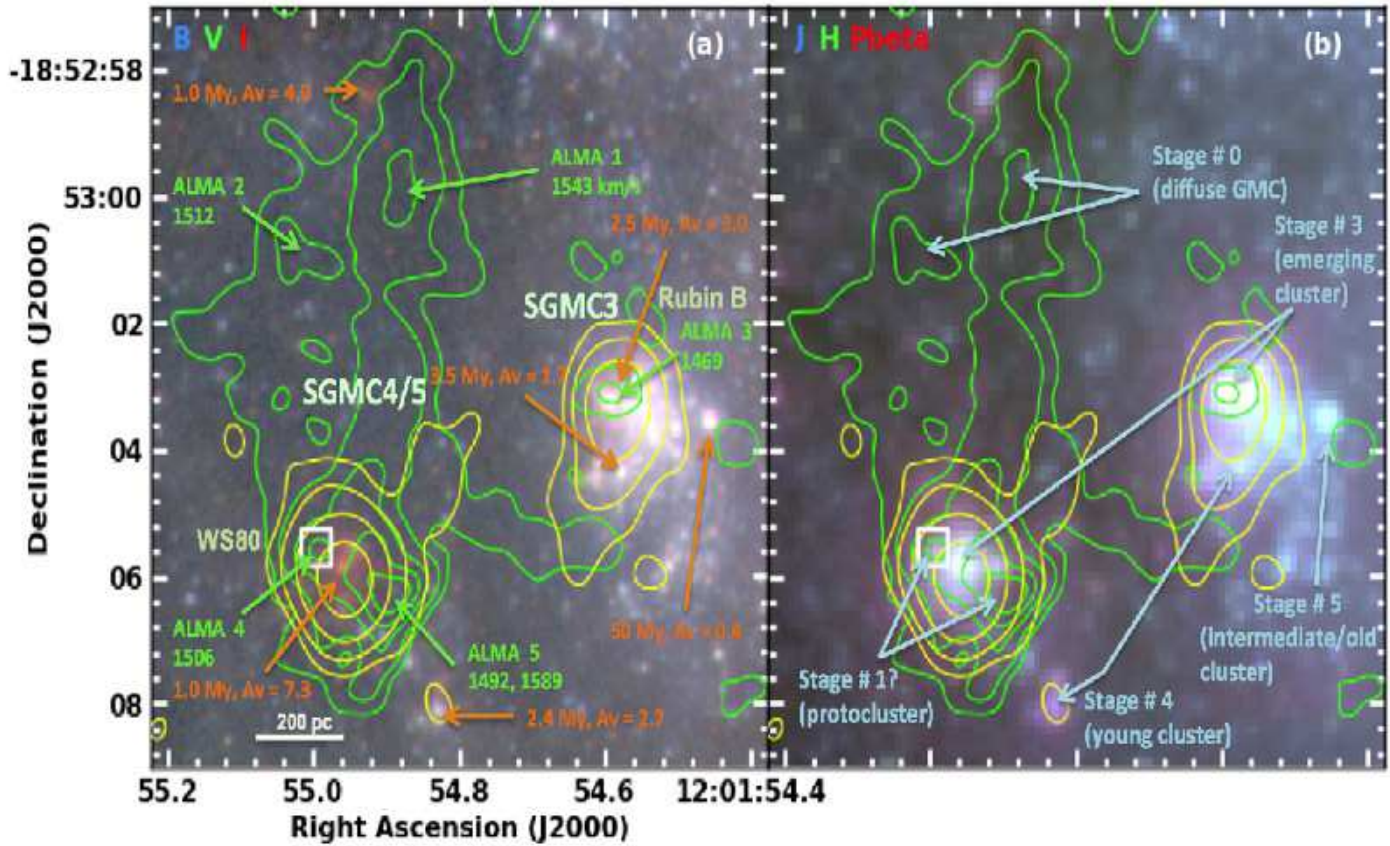}  
\end{center}
\caption{Correspondence between CO knots (green), 3.6~cm radio emission (yellow) and optical star clusters (orange) in SGMC3 (including Rubin~B) and SGMC4/5 (including WS80; see Whitmore and Zhang 2002 for a discussion).  The green contours (70, 170, 370, 570, 770~K~km~s$^{-1}$) show CO emission; yellow contours (60, 180, 540, 
1640~$\mu$Jy beam$^{-1}$)  show radio emission. See Figure~7 for a more detailed description of the figure. 
}
\end{figure}
\end{landscape}

\begin{landscape}
\begin{figure}
\begin{center}
\includegraphics[width = 4.5in, angle= -90]{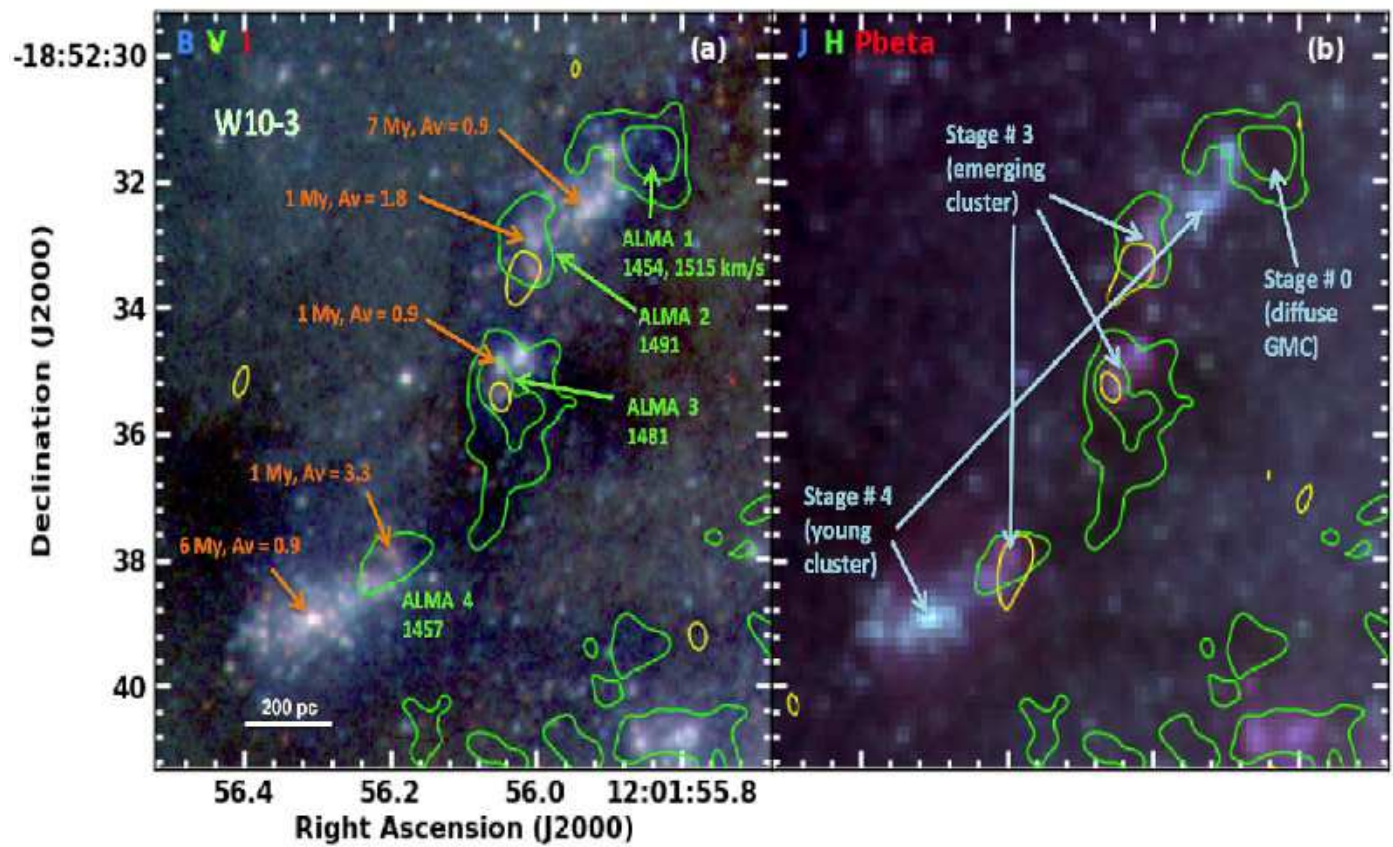}  
\end{center}
\caption{Correspondence between CO knots (green), 3.6~cm radio emission (yellow) and optical star clusters (orange) in W10-3 (see Whitmore et~al.\  2010 for location of W10-3). The green contours (10, 30~K~km~s$^{-1}$) show CO emission; yellow contours (40~$\mu$Jy beam$^{-1}$)  show radio emission. See Figure~7 for a more detailed description of the figure. 
}
\end{figure}
\end{landscape}

\begin{landscape}
\begin{figure}
\begin{center}
\includegraphics[width = 5.0in, angle= -90]{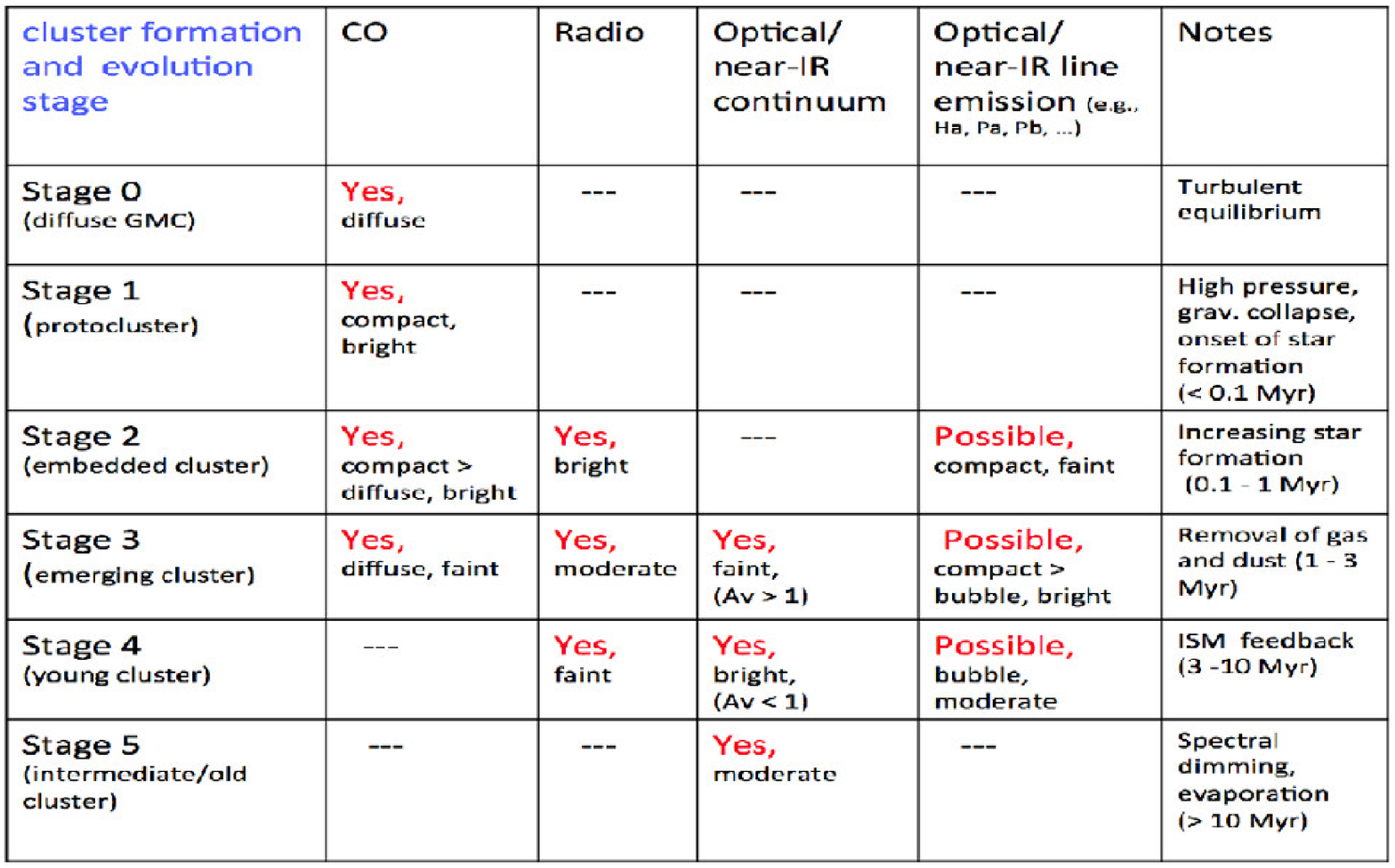}  
\end{center}
\caption{Matrix showing the definition of the various stages of cluster formation and evolution used in the classification framework outlined in \S5.}
\end{figure}
\end{landscape}

\begin{figure}
\begin{center}
\includegraphics[width = 5.25in, angle= 0]{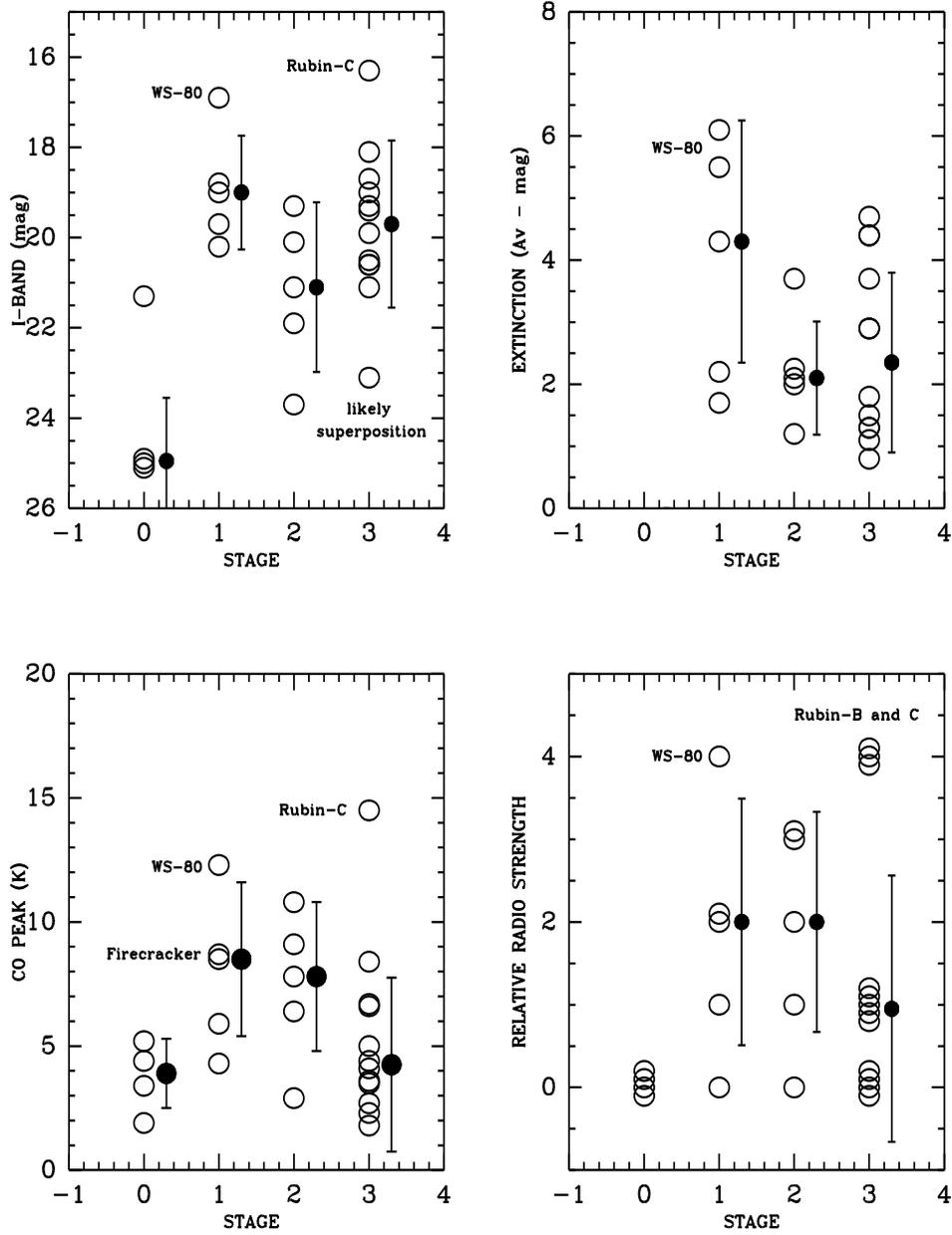}  
\end{center}
\caption{Plots of evolutionary stage vs.\ CO peak, relative radio strength (max), total $I$-band magnitude for clusters (corrected for extinction), and extinction ($A_V$) from Table~1. For regions without clusters, the 
$I$-band magnitude is set at 25 in order to include a value on the plot. For the radio strength, which is defined in \S6,  an incremental value
of 0.1 has been added to redundant points in order to make them visible on the figure. The solid circles show the median values and 1$\sigma$ uncertainties for each stage and are offset slightly to make them more apparent. See Notes to Table~1 for more details. }
\end{figure}

\begin{deluxetable}{lccccrccccl}
\tabletypesize{\footnotesize}
\rotate
\tablewidth{0pt}
\tablecolumns{11}
\tablecaption{Multi-Wavelength Data for Objects Discussed in this Paper\label{tab:alpha}}

\tablehead{
\colhead{Region Name (aka)}  & \colhead{R.A.\tablenotemark{a}} & \colhead{Dec.\tablenotemark{a}} & \colhead{CO Flux\tablenotemark{a}} & \colhead{Velocity\tablenotemark{a}}  & \colhead{Radio Strength}  & \colhead{\# cl} & \colhead{Total I} &  \colhead{Age}  & \colhead{$A_V$}    
& \colhead{Stage}\\
&  (J2000) &  (J2000) & (K) & (km s$^{-1}$) &  & & (mag) & (Myr) & (mag) 
}

\startdata
LT-ALMA-1 &      180.47858  & $-$18.87320  &  4.3    &   1606  & \multicolumn{1}{c}{---} & 5\rlap{\tablenotemark{b}} & 20.7  & 3.2 & 1.7  & 1 (?)\tablenotemark{c} \\
LT-ALMA-2 & 180.47865 & $-$18.87382       &  2.9     &   1610 & \multicolumn{1}{c}{---}  & ---  &  --- & --- & ---  & 2 \\
LT-ALMA-3 &   180.47894 &	$-$18.87476   &  3.5    &   1597  & \multicolumn{1}{c}{---}  & 1  & 23.3  & 1.0   & 3.7  & 3  \\
LT-ALMA-4 & 180.47887 &	$-$18.87538     &   4.1    & 1602    &\multicolumn{1}{c}{---}   & 1 & 23.2  & 1.0 & 4.4  & 3  \\
LT-ALMA-5 &  180.47896 &	$-$18.87595 & 1.8     & 1606     &  \multicolumn{1}{c}{---} & 6  & 19.9  & 4.8 & 1.1  & 3   \\
LT-ALMA-6 &  180.47977 &	$-$18.87718    & 4.4     &  1597   & \multicolumn{1}{c}{---}  & 1  & 23.4  & 3.3 & 4.7 & 3  \\
LT-ALMA-7 &  180.47946 &	$-$18.87762    &  5.0    &  1585   & weak (1)  & 5  & 21.3  & 1.0 & 4.4 & 3     \\
LT-ALMA-8 &   180.47990 &	$-$18.87798   &  6.4      &  1574   & moderate (2)  & 1  & 23.3  & 1.0 & 3.7 & 2    \\
\\

SGMC1-ALMA-1 (Rubin C/D)  &  180.48140  & $-$18.87930     & 8.5     & 1385    & weak (1) &  7\rlap{\tablenotemark{b}} & 20.3  & 1.7 & 2.2  &  1 (?)\tablenotemark{c} \\
SGMC1-ALMA-2 &  180.48077 & $-$18.87997    &  \llap{1}0.8    & 1380    & strong (3) & 4  & 20.8  &  1.0 & 1.2 & 2     \\
SGMC1-ALMA-3 (Rubin C) & 180.48062  & $-$18.88025  &\llap{1}4.5   &   1424  & very strong (4) & 8 & 18.9  & 1.0 & 1.3  &  3 \\
SGMC1-ALMA-4 & 180.48032 &  $-$18.88010     &  9.1    & 1375    & strong (3)  & 6 & 20.5 & 1.0 & 2.0  & 2   \\
SGMC1-ALMA-5 &   180.48035 & $-$18.88067   &   4.4   & 1510     &  \multicolumn{1}{c}{---} & 4  & 21.3  & 4.2 & 0.1 &  0   \\
\\

SGMC2-ALMA-1 & 180.47820 & $-$18.88087     & 6.6   & 1421    & very strong (4)  & 3 & 20.7 & 2.5 & 2.9  & 3 (?)\tablenotemark{d}  \\
SGMC2-ALMA-2 &   180.47867 & $-$18.88107   &  7.8     & 1450    & weak (1) & 2  & 23.1  & 3.0 & 2.1 & 2  (?)\tablenotemark{d}  \\
SGMC2-ALMA-3 &  180.47878 & $-$18.88130     & 8.4      &  1467   & weak (1)  & 1 & 23.6  & 130 & 0.8  & 3 (?)\tablenotemark{d} \\
SGMC2-ALMA-4 (Firecracker)&  180.47807 & $-$18.88143    &  8.7    & 1519    & moderate (2)  & 3\rlap{\tablenotemark{b}} & 21.3  & 2.5 & 4.3 & 1   \\
\\

SGMC3/4/5-ALMA-1 & 180.47870 & $-$18.88333  	&   5.2    &  1543   &  \multicolumn{1}{c}{---}  & ---  & --- & --- & ---  &  0 \\
SGMC3/4/5-ALMA-2 &   180.47932 & $-$18.88356	&    3.4   &  1512   & \multicolumn{1}{c}{---}   & ---  &---  & ---  & --- &   0  \\
SGMC3/4/5-ALMA-3 (Rubin B) &  180.47754  & $-$18.88420 	&   6.7    &  1469   & very strong (4)  &  8 & 18.0  & 1.0  & 2.9 &   3  \\
SGMC3/4/5-ALMA-4 (WS-80) &180.47914 &$-$18.88492 
&\llap{1}2.3   & 1506    & very strong (4)  & 2  & 20.5  & 1.0  & 6.1 
& 1 (?)\tablenotemark{c} \\
SGMC3/4/5-ALMA-5 (WS-80) &   180.47868 &  $-$18.88510	&  5.9     &   1589  & moderate (2)   &  1 & 23.4 & 2.9 & 5.5 & 1 (?)\tablenotemark{c} \\
\\
\tablebreak
W10-3-ALMA-1 &   180.48264 &	$-$18.87538   &   1.9   &  1515   & \multicolumn{1}{c}{---}  & ---  & ---  & --- & --- & 0   \\
W10-3-ALMA-2 &   180.48337 & 	$-$18.87582   & 2.7     &  1491   & weak (1)  &  5 & 21.0 & 1.0 & 1.8 &  3   \\
W10-3-ALMA-3 &  180.48352 &	$-$18.87643    & 3.6      & 1481    & weak (1) & 3  & 20.2   & 1.0 & 1.3 &  3   \\
W10-3-ALMA-4 &   180.48413 &	$-$18.87724  & 2.3     & 1457    & weak (1)  & 4 & 21.4 & 3.1 &  1.5  & 3 \\
\enddata

\tablecomments{
column~1~= Name of region as defined in Figures~7 through~11, 
column~2~= Right Ascension (J2000),
column~3~= Declination (J2000),
column~4~= CO (3-2) peak values from CPROPS (Rosolowsky \& Leroy 2006),
column~5~= CO (3-2) velocity (km~s$^{-1}$) from CPROPS,
column~6~= estimated maximum radio (3.6~cm) strength 
(weak~$=1<100\,\mu$Jy beam$^{-1}$, moderate~= 2 -- 100 to 
200\,$\mu$Jy beam$^{-1}$, strong~= 3 -- 200--400\,$\mu$Jy beam$^{-1}$, very strong~= 4 -- $400\,\mu$Jy~beam$^{-1}$), based on visual estimate from contours in Figure~7 through~11, column~7~=  number of clusters within 24~pixel ($1.2''$ -- 130~pc) box and $I$-band magnitudes brighter  than 24 (from Whitmore et~al.\ 2010), column~8~= total $I$-band magnitudes (uncorrected for extinction) for all clusters  in 24~pixel box (from Whitmore et~al.\ 2010), column~9~= mean age for clusters in Myr (from Whitmore et~al.\ 2010), column~10~=  mean value of extinction 
($A_V$) for clusters (from Whitmore et~al.\ 2010), column~11~= estimated  evolutionary stage from classification system defined in Figure~12 and
 \S5.  }
\tablenotetext{a}{Based on preliminary catalog from Leroy et~al.\ 2014 (i.e., Paper 2).}
\tablenotetext{b}{The physical association of clusters with these regions is uncertain since they are  on the edge of the region. }
\tablenotetext{c}{Classification is uncertain since: 1)~threshold may not be sufficient to show presence of radio emission, or 2)~radio emission is present along the line-of-sight but may be associated with a nearby object, as appears to be the case with SGMC2-ALMA-4 (see Johnson et~al.\ 2014) for example. }
\tablenotetext{d}{Classification is uncertain due to possible superposition (see discussion in \S6.3).}
\end{deluxetable}

\end{document}